# A systematic literature review on process model testing: Approaches, challenges, and research directions

Kristof Böhmer[*], Stefanie Rinderle-Ma[*]

*University of Vienna, Faculty of Computer Science, Research Group Workflow Systems and Technology, Währingerstrasse 29, 1090 Vienna, Austria*

**Abstract**

*Context:* Testing is a key concern when developing process-oriented solutions as it supports modeling experts who have to deal with increasingly complex models and scenarios such as cross-organizational processes. However, the complexity of the research landscape and the diverse set of approaches and goals impedes the analysis and advancement of research and the identification of promising research areas, challenges, and research directions. Hence, a comprehensive and systematic review that covers state of the art research in the process model testing domain is required.

*Objective:* The objective is to analyze process model testing approaches to identify and present, connected research domains and the used terminology. Additionally, research challenges, gaps, and future research directions are identified and discussed.

*Method:* A systematic literature review is conducted to identify interesting areas for future research and to provide an overview of existing work. Over 6300 potentially matching publications were determined during the search (literature databases, selected conferences/journals, and snowballing). Finally, 153 publications from 2002 to 2013 were selected, analyzed, and classified.

*Results:* It was found that the software engineering domain has influenced process model testing approaches (e.g., regarding terminology and concepts), but recent publications are presenting independent approaches. Additionally, historical data sources are not exploited to their full potential and current testing related publications frequently contain evaluations of relatively weak quality. Overall, the publication landscape is unevenly distributed so that over 31 publications concentrate on test-case generation but only 4 publications conduct performance test. Hence, the full potential of such insufficiently covered testing areas is not exploited.

*Conclusion:* This systematic review provides a comprehensive overview of the interdisciplinary topic of process model testing. Several open research questions are identified, for example, how to apply testing to cross-organizational or legacy processes and how to adequately include users into the testing methods.

---

[*]Corresponding author. Tel.: +43-1-4277-79125; Fax: +43-1-4277-89125
*Email addresses:* `kristof.boehmer@univie.ac.at` (Kristof Böhmer),
`stefanie.rinderle-ma@univie.ac.at` (Stefanie Rinderle-Ma)

## Contents

## 1. Introduction

Over the past years, processes have started to be deeply integrated in many domains such as business, scientific applications, and e-learning [1]. Hence, ensuring the stability, correctness, and feature completeness of processes has become a crucial challenge and an emerging major research topic.

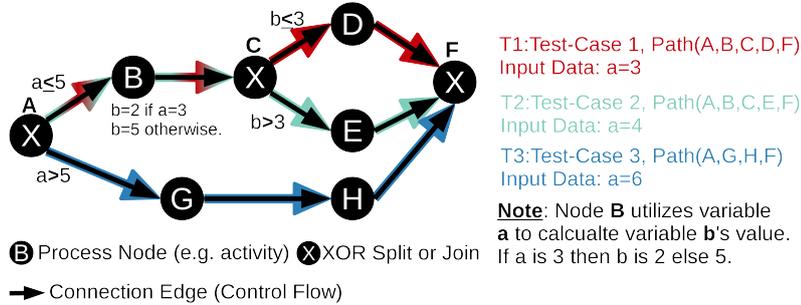

Figure 1: Process model with three example test-cases.

*Terminology.* Processes have risen to deeply integrated solutions which are extremely important for various organizations [137]. During this development, multiple definitions emerged [2]. For this Systematic Literature Review (SLR) a process is defined as '*a structured set of activities designed to produce a specified output for a particular customer or market*' [3, p. 5].

In addition, this SLR concentrates on process testing (i.e., testing of processes models for properties such as conformance, correctness, or quality of service using test-cases) approaches. Hence, testing has been defined as '*operating a system or component under specified conditions, observing or recording the results and making an evaluation*' [4, p. 76].

Figure 1 illustrates how one of many possible types of test-case definitions can be used to ensure that process model activities have an adequate (i.e., correct) internal behavior. In this example, the test-cases consists of an expected *execution path* which should be followed by the process model during the test-case execution and input data that should guide (e.g., at conditions) a correct process model definition to follow the predefined expected execution path (cf., [S27]). During the test-case execution an instance of the process model is created and instrumented using the test-case input data. Subsequently the process model instance execution is monitored and the executed process model path is compared to the expected execution path. A fault can, for example, be identified if a process instance does not follow the expected execution path or if an activity reports an internal error (e.g., if T1 would 'visit' Node E because there was an error at Node B which did set variable b to some value greater than 3).

*Research challenges.* **Cross-Organizational Scenarios**: A typical automobile manufacturer, for example, uses thousands of processes in different domains such as research, production, or management. Hence, current processes frequently integrate a diverse set of resources and external partners, thus creating *cross-organizational processes* [5]. At such a complex process a single undetected error can result in lawsuits or loss of confidence [6]. Therefore, it becomes necessary to constantly apply *tests* to ensure, for example, process soundness and compliance [7]. However, testing cross-organizational process models is substantially more complex than testing, for example, process models which are only used internally at one organization. Why? Because external partners can change noticeably (e.g., until some user reports an error that a process no longer executes correctly). Additionally the partners must frequently be handled as a black boxes, so it is not possible to observe their internal behavior which would be helpful to check their correctness. Finally, the respective partners can be chosen dynamically just in time during the process model execution which makes it harder to create deterministic tests. Additionally the testers have to deal with network communication, latency, service level agreements, and typically a more diverse error handling mechanic which must be tested throughly. Note, complex error handling is often specified because, for example, there can be timeouts, corrupted data, or unspecified partner behavior which must be 'foreseen' and handled by exception paths. The needs for cross organizational processes also exist in other domains such as education, science, or health care.

**Volatile Environments**: Process models must constantly be adapted to react on changing environmental conditions and requirements [8]; a great challenge for companies/organizations and their employees [137]. However, by applying *tests* it can become easier to ensure quality and stability [S14] because, testing allows to identify unintentional errors and no longer executable use-cases (e.g., caused by a lack of employee skills or training [9]). This applies especially to *long used processes repositories* where the original process designers frequently left the organization, documentation is outdated or non-existent, and a constant time pressure requires to quickly implement adaptations into feature crowded process models (*legacy processes*).

**Flexibility and Costs**: Process testing approaches also struggle with additional challenges. One of them is related to the flexibility of process modeling languages, which allow to integrate various components [6]. Hence, testing approaches have to deal with the simulation or integration of users, automatic systems, and various partners [S65,S97]. For example, it's important that the tests do not increase interaction liabilities (payed for accessing external partners) to an unjustifiable extend (cf. [S12] and [10]).

**Diversity**: Another challenge is the diversity of up-to-date process modeling languages [11]. For example, the WS-Business Process Execution Language (BPEL) [12] depends on specific technology (e.g., XPATH). Others, such as the Business Process Model and Notation (BPMN) [13] describe a similar process behavior using more abstraction and a technology independent approach. Testing approaches therefore need to keep the specific characteristics of process modeling languages and their enactment systems in mind to find the right balance between flexibility and specialization [S92].

These are just some examples which demonstrate the flexibility and complexity of the 'process world'. However, testing is a promising approach to address increasingly complex process scenarios and can also be applied complementary to existing approaches which concentrate mainly on soundness and compliance

(e.g., [14,15]).

*Motivation for the systematic literature review.* Testing is used at the software development domain since more than 20 years [16]. Hence, we assume that the process testing community is influenced by the software domain. However, it has not yet been investigated if an interlocking exists. Furthermore, the process modeling community has already created conformance checking (e.g., [17]) or validation (e.g., [18]) techniques which are not based on tests. Hence, the question arises which challenges are addressed by process testing approaches (e.g., to asses the value of the test based approaches or to identify not sufficiently covered areas). In this context the question arises if current process testing research is already covering upcoming challenges. For example, cross-organizational process scenarios grew in importance over the past years; are those already supported by process testing approaches? Subsequently, it's also important that potential users are capable of applying the presented approaches (e.g., which skill-set is needed? Was the applicability of current approaches proven through user and case studies?). Finally, process testing emerged to a complex and extensive research domain. It is therefore hard to get an overview above the existing work. Hence, a SLR would greatly contribute to the understanding and future development of the process testing discipline.

This paper is organized as follows. Section 2 starts by outlining the applied research methodology. Hence, it discusses the research questions, literature search (6256 papers were discovered), literature selection (153 relevant papers were identified), data extraction and classification strategy. The results are presented in Section 3, which is also discussing existing SLRs and identifies the main testing approaches. Classification techniques are utilized in Section 4 to compare the identified *main testing approaches* with the *required skills/ knowledge* and to analyze if existing test approaches support *upcoming challenges*. The identified gaps and promising challenges are discussed in Section 5. Subsequently, Section 6 discusses the limitations of this SLR. The paper is concluded in Section 7.

## 2. Research methodology

This SLR is based on a systematic literature review (cf. [19–22]). Multiple guidelines and best practice recommendations were investigated, ranging from domains like psychology [23] to software development [21]. To our knowledge there are currently no guidelines available which specifically focus on the needs and requirements existing in the process modeling domain. Hence, we examined existing reviews and guidelines (cf. [19–21,24]) to form an appropriate approach.

The applied research methodology is condensed into Figure 2. First the need for this review is identified (cf. Section 1) by investigating multiple papers (including a preliminary search) and discussions with experienced researchers. Those preparatory steps allowed us to create the initial results which we considered to be analyzable by a SLR. Then we continued by searching/checking already existing SLR in the process modeling domain. Subsequently, the authors designed the SLR and documented it within a *review protocol*. This was followed by a literature search, data preparation, and date extraction. synthesis into multiple categories. Finally, the results were documented, classified, and used to identify novel research challenges and research areas.

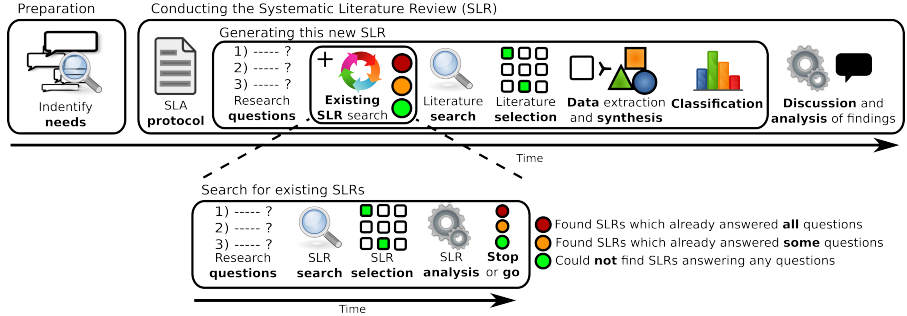

Figure 2: Research methodology.

2.1. Research questions

Research Questions (RQ) are extremely important for any systematic review [22]. Hence, we start by discussing our RQ.

**RQ1** What does testing in process model design related areas mean (terminology, common understanding, related research areas)?

**RQ2** Which testing approaches are currently in use to test process models and how are their objectives defined?

**RQ3** What are the main process testing areas the research community is currently concentrating on?

**RQ4** How are up to date testing approaches suited for current process challenges?

**RQ4.1** Which skills/knowledge and manual work are required to prepare a test-case?

**RQ4.2** How are current approaches dealing with upcoming challenges such as legacy processes or cross-organizational scenarios?

RQ1 to RQ3 focus on generating an understanding of the complex topic and its diverse research directions, while RQ4, RQ4.1, and RQ4.2 focus on identifying the suitability of the published approaches for pressing challenges in the process modeling and testing domain.

RQ1 concentrates on finding out how testing is defined by the research community and which research areas are interconnected. This definition of RQ1 enables the identification of relevant resources by building up keywords to search for, forming search strings, and by identifying relevant sources.

The search results are then analyzed in order to identify the main approaches and objectives of up to date process testing research to address RQ2 and RQ3.

Classification approaches are considered by the authors as an promising approach to address RQ4.1 and RQ4.2. For RQ4.1 a classification approach will be used that compares the required knowledge/skills/training (to utilize the available testing approaches) with the most important testing techniques (e.g., unit or regression testing). For RQ4.2 it will be analyzed if current testing approaches were designed to support legacy processes or cross-organizational scenarios. Hence, RQ4 aims on finding research challenges and gaps [22].

It's generally accepted as a best practice approach to check if a new SLR is really necessary upfront before conducting it [22]. Hence, we are adding a preliminary RQ0 which specifically focuses on determining if RQ1 to RQ4 (including their sub-questions) can be answered by existing SLRs.

**RQ0** Are existing systematic literature reviews already providing answers for the formulated research questions (RQ1 to RQ4) in whole or in part?

*2.2. Literature search*

The literature search was realized in a three-step process which is executed twice, once for the search for existing SLRs and, afterwards, once for the main SLR to identify relevant research papers which answer the discussed research questions.

For the initial search for existing SLRs we concentrated our efforts on large digital research databases, especially SCOPUS, IEEE Xplore, DBLP, and ISI Web Of Knowledge (includes the Springer database). Due to the broad base of sources we assume that no existing SLR was left behind.

The main SLR applies a more targeted approach which utilizes a combined vertical and horizontal search. The horizontal search is based on digital research databases (SCOPUS, ISI Web Of Knowledge, and ACM Digital Library), followed by a manual vertical searches which focuses on selected top ranked sources. A detailed list can be found in the appendix.

Different search terms were used for searching existing SLRs (a) and for searching interesting papers for this new SLR (b). We used a systematic method to construct the necessary terms, adapted from [23]. For a) the search string was built by combining terms (using OR/AND) which identify SLRs (literature review, research review, systematic review, research synthesis, meta-analysis), interesting areas (validation, verification, test, conformance, compliance, constraint), and terms which narrow the results on processes (workflow - workflow is frequently used as a synonym for process, therefore, we added it to the search keywords list -, business process). For b) the search string was composed of the interesting area (test) and narrowing terms (workflow, business process). Stemming and wildcards (e.g., test* to cover also testing) were applied where possible.

The time frame for the search for primary papers was limited to papers which were released at the year 2013 or earlier. This strengthens the reproducibility of this SLR because the literature search was conducted during the second and third quarters of 2014. Hence, all papers from 2014 were not already released and researchers would probably gain different search results if they are reproducing this SLR, for example, in 2015. The search for existing SLRs was not limited.

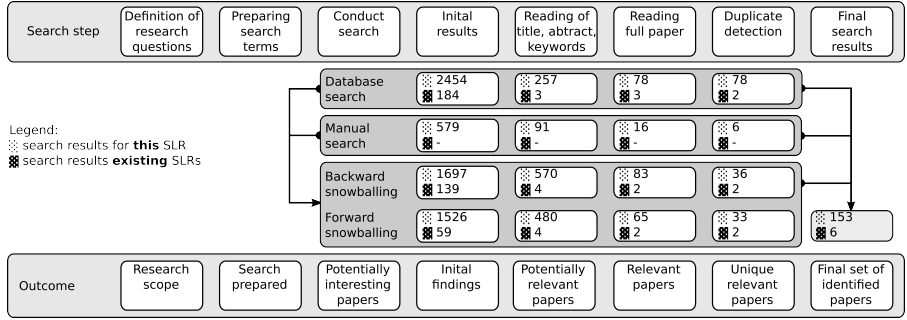

Figure 3: Overview of the (intermediate) search results.

After the initial search, backward and forward snowballing [25] were applied.

The search for existing SLRs identified 382 papers. 6256 papers were identified during the search for the new SLR (cf. Figure 3). Filtering for duplicates and applying inclusion/exclusion criteria to assure the relevance of each individual paper resulted in following findings: 6 relevant SLRs exist in a similar area (from 382 papers) along with 153 papers (from 6256) that were discovered as relevant for this new SLR.

2.3. Literature selection

Following on from the search, the raw data was aggregated. Subsequently, the SLR process was continued by filtering the papers based on criteria which were defined and documented upfront at the SLR protocol. The documented selection/filtering process consists of *two* main steps, whereby each of those contains two to three sub-activities (cf. Figure 3). First, the paper's *title* and *abstract* were checked. Second, the *introduction*, *conclusion*, and (if the authors still could not clearly identify it as relevant or not relevant) the *full text* were read and checked if the paper is related to processes and testing.

This SLR applied the following inclusion and exclusion criteria. To be included a paper must a) be a primary paper (e.g., no doctoral theses – it's expected that the findings are also published as a paper), b) focus on testing (for this new SLR and existing SLRs) and related areas (existing SLRs only), and c) processes. The d) abstract, title, or keywords must contain relevant terms such as *business process* or *test* (compare Section 2.2). In addition, the paper must e) be published in English (enabling international researchers to reproduce the findings) and the publication must f) be published in a peer reviewed journal or conference. The exclusion criteria excluded a) short/poster papers (the existence of a full length paper was assumed), papers whose b) full text was not accessible, and papers that are c) not presenting novel approaches (e.g. papers that only present a summary of existing approaches or duplicates).

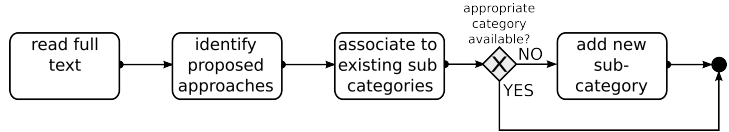

Figure 4: Data extraction and aggregation: categorization approach

A paper was immediately excluded if at least one of the exclusion criteria could be matched. In addition, it was required that a paper matches all inclusion criteria to pass the selection process. An extended duplicate removal was applied, for example, if a paper was published at a conference and at a journal then the journal paper was preferred.

The relevance criteria for existing SLRs that potentially contain answers for the given research questions used a wider focus to assess related areas such as conformance checking, verification, or validation (cf. Section 2.2). This decision was made to reduce the likelihood that some relevant publication will be overlooked.

Altogether, by applying these criteria it was possible to significantly reduce the number of papers from 6638 to 159 (cf. Figure 3). These 159 papers were processed in the following.

*2.4. Data extraction and aggregation*

This SLR extracts information such as the paper's main topic of interest, the non-functional and functional objectives of the presented testing approaches, and the applied validation and evaluation techniques. A multitude of sub-categories could be assigned to each of those top categories, for example, the paper's main topic could be categorized as regression or performance testing. Hence, we defined a manual extraction process (cf. Figure 4) which started with a base set of sub-categories (e.g., inspired by existing literature, such as [S58]). In the following the list of sub-categories was extended when necessary. Finally, the categories were analyzed and aggregated to form a coherent whole using a constant granularity level.

The extracted information can be separated into three groups. The first one contains details mainly used for documentation and organization purposes (e.g., a unique id), the second one contains generic paper information (e.g., author names), the third one contains information such as the abstract, a short summary of the main ideas and concepts, and boolean values representing if the authors, for example, address cross-organizational scenarios.

However, the vast majority of extracted information was supported by the described manual categorization approach, namely the paper's main objective (e.g., to present an unit testing approach), functional and non-functional test objectives, required knowledge/skills to prepare a test, the process life cycle where it can be used, which validation/evaluation data and approaches are used (if any), which process views are tested (e.g., control or data flow), the stakeholders which are addressed by the approach (e.g., domain experts), and the inspiration source of the paper (e.g., if the presented approach was ported from the software development domain to the process testing domain). Altogether 28 individual values were extracted from each paper.

## 3. Results and findings

First, we will outline which relevant existing systematic reviews were discovered and if those already provide answers for our research questions. This step will be followed by outlining publication dates of the identified primary papers.

In the following, RQ1 will be tackled by a general discussion what testing in process design related areas stands for (see Section 3.2). This section is continued by an overview about the identified main testing areas addressed by existing publications (e.g., test-case generation, regression testing, and so on; cf. Section 3.3) to address RQ2 and RQ3.

### 3.1. General overview on relevant publications

This section starts by outlining the results of the search for existing relevant SLRs. Afterwards temporal information of the selected relevant publications will be presented.

#### 3.1.1. Existing systematic literature reviews

By searching literature databases, and applying snowballing to widen the search area, six unique, relevant (out of 382 search results) publications were identified which matched the following criteria: the paper must a be a SLR, focusing on process modeling and test related areas. Subsequently, each SLR was read in full text, analyzed and checked if its research questions or findings were already answering one or more research questions of this new SLR – to address RQ0.

However, it was found that existing SLRs are not able to completely answer the research questions of this SLR. On reason for this finding was that existing SLRs [S58–S63] only focus on very narrow areas and, therefore, provide only an incomplete picture which makes it impossible to asses the research directions and challenges of the whole process testing domain (e.g., some process testing areas are not even covered by a single SLR). [S61], for example, concentrates only on single testing approach (unit testing) and BPEL processes. Hence, it leaves other testing approaches (e.g., regression testing) and modeling languages (e.g., BPMN or Petri Nets) behind. Additionally, it was recognized that the existing SLRs were, typically published around the year 2010 or earlier. Hence, they are missing the latest developments. Moreover, the identified SLRs frequently only concentrated on web service related technologies that represent only a small excerpt from the whole process modeling domain. For this reasons it was concluded to continue with performing this SLR.

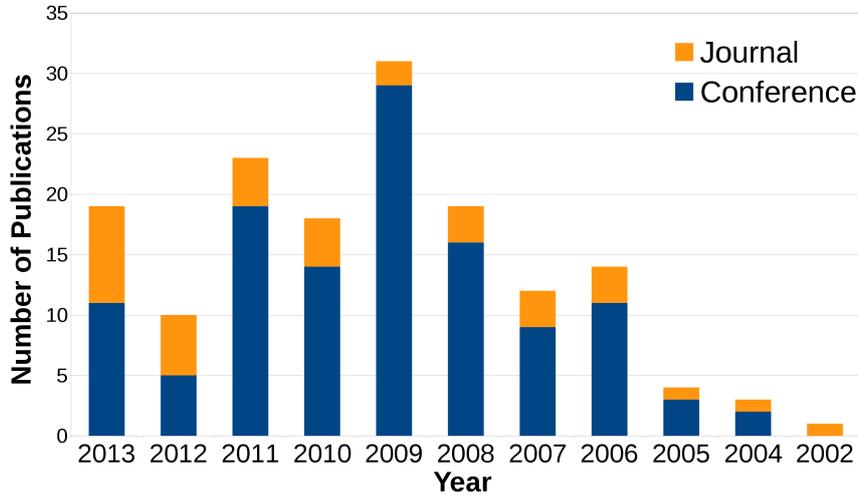

Figure 5: Publications on process model testing published from 2002 to 2013.

*3.1.2. Publication time frame*

The literature search discovered 153 relevant papers (released between 2002 and 2013[1]). A detailed overview can be found in Figure 5 which shows that the interest in this topic has constantly increased over the years. During the early years (2002-2005) there were only 0 to 4 publications per year. From 2006 to 2008, the publications increased to 12-19 papers each year. Beyond 2009, 10-31 publications were released each year concentrating on mature topics like cross-organizational scenarios or test data generation. In 2009, 31 papers were released. Most likely this is related to the release of BPEL 2.0 in 2007 (cf. [S6]).

This development indicates that the scientific community has a high interest in the process model testing domain. research.

*3.2. Testing in the process domain*

This section addresses RQ1. Our findings indicate that a common inspiration for the identified publications is the software development domain. There, BPEL is frequently used when developing service oriented software. Hence, authors which already published on software testing then adapted known concepts and ported these to the process testing domain (e.g., [S109]). This also influenced the terminology so that terms like regression or unit testing can be found in both domains.

---

[1]Papers published in 2014 are not included in this SLR to strengthen the reproducibility of this SLR (cf. Section 2.2).

*Terminology, definition, and common understanding.* Several relevant definitions, used in the testing domain, are currently available and published in standards by the ISO or IEEE. For example, the IEEE 1012-2012 defines test-cases as a '*set of test inputs, execution conditions, and expected results developed for a particular objective*' [26, p. 11]. The definition already states that a test-case must consist of the required test data, has to set the execution conditions so that the test can be executed, and that it must define the properties which should be checked by the test-case. Hence, it divides the test-cases in three individual problem domains which can be tackled separately. This is also reflected by research where publications are frequently only focusing one specific challenge such as the test data generation. Those test-cases are then executed during the so called *testing* which was defined, at the IEEE 829-2008, as an '*activity in which a system or component is executed under specified conditions, the results are observed or recorded, and an evaluation is made of some aspect*' [27, p. 11].

This exemplary definitions provide already the insight that, in general, testing consists of specifying and checking if specific characteristics are met by the system.

Nevertheless, it must be taken into account that those definitions are tailored towards requirements of the software engineering domain which has a different basis for the utilization of tests than the process modeling domain. For example, the granularity is different. Even small software components frequently contain way more than a thousand lines of code. A process model, instead, which contains a thousand steps would be recognized as a large enterprise solution and each step would provide much more 'value' compared to a single line of code. Similar diversions become visible when the utilized log generation practices are compared between software development and process modeling. Process executions are frequently logged extensively, sometimes the whole data and control flow can be extracted from the log (cf. [28]). A comparable extensive logging approach (e.g., one log output after each method call or changed variable) is typically not applied at the software development domain [29]. The two domains (software testing vs. process model testing) also struggle with different challenges. For example, testing concurrent executions is a major challenge at the software testing domain [30] while on the process model testing domain it is 'solved' by multiple authors, for example, by generating a unique test-case for each possible process model execution path. This is possible, because process models typically contain way less possible execution paths and individual nodes (i.e., statements in software testing) than source code. Hence, one of the most serious challenge when testing concurrent executions, potential *state explosions* (i.e., more test-cases/potential execution paths must be checked than possible in a reasonable amount of time), is mitigated through relatively simple process model structures and various state explosions prevention techniques originating from formal testing approaches [S33]. Hence, generating and testing all potential concurrent node execution orders and execution paths is frequently possible, cf. [S19,S33,S113,S123].

The differences, mentioned above, have created a diverse field of techniques and concepts that use unique approaches (not based on testing) to tackle prob-lems like validation, conformance checking or verification at the process domain. Conformance checking, for example, analyses if a model's implementation/execution mimics the process model or fulfills specific requirements. This objective can be achieved by tests [S81,S91] or by a different technology called log mining [31,32]. Another example are soundness properties like deadlocks or livelocks and their validation. One possibility is to transform the process model into a formal representation (e.g., a Petri Net) and various tools (e.g., the CPN-tools) and algorithms can then be applied (cf. [33]). Of course, a check for soundness properties can also be implemented with tests. For example, by checking if all tests have terminated (for deadlock detection) or by applying a static analysis of the process's control flow (by e.g., testing for known deadlock patterns [S66]).

All those alternatives slowed down the application of tests at the process research community, because there were already existing specialized concepts available which could be pushed forward. So, the initial testing based solutions were mainly published by researchers with a strong software development background. But today a increasing interest (cf. Section 3.1.2) into this topic has emerged at the process modeling research community. Therefore, novel testing concepts are developed that fulfill upcoming challenges of the process modeling domain, which are discussed in the following paragraph.

*Related research areas.* We found that existing publications are relying on already existing software testing definitions and standards (e.g., what unit testing stands for). [S32], for example, express that processes have '*built-in constructs expressing concurrency and synchronization, thus can be seen as a kind of concurrent programs.*'[S32, p. 75].

Nevertheless, there are also some challenges arising independently of the software testing domain. This includes testing in cross-organizational scenarios which is perceived as '*complicated substantially by heterogeneous process-modeling languages*' [34, p. 59] or the testing of legacy processes (cf. Section 5). Additionally, inter-organizational processes are seen as a key challenge because of their loose coupling, heterogeneity in modeling languages and distributed partners [34] for process testing. Another key challenge '*in testing a composite service emerges from its inherent variability*' [S56, p. 138] and the demand for approaches which allow to describe and test the semantic characteristics of a process [35].

### 3.3. Testing of processes and workflows

In this section, RQ2 and RQ3 are examined by investigating current testing approaches and techniques in the process modeling domain. During the data extraction process (cf. Section 2.4) each publication was clustered and sorted based on their main topic/objective to form groups of related publications. These groups were created based on the number of papers covering one specific objective (so if more papers concentrate on an objective then it's more likely that this objective is used as a group) and their delimitation to the objectives of other groups. Performance testing publications, for example, although very few (4) were discovered, aim to fulfill so distinctive objectives that they cannot be appropriately merged with another group (cf. Table 1).

| Title | No. |
|---|---|
| Unit testing (Section 3.3.1) | 14 |
| Test-case generation (Section 3.3.2) | 56 |
| Test data generation (Section 3.3.3) | 36 |
| Integration testing (Section 3.3.4) | 7 |
| Quality of Service testing (Section 3.3.5) | 19 |
| Conformance testing (Section 3.3.6) | 21 |
| Regression testing (Section 3.3.7) | 13 |
| Formal testing (Section 3.3.8) | 28 |

Table 1: Identified main objectives. Note, a publication that concentrates on multiple topics and will be counted multiple times.

Each of the following sections is addressing one main testing approach (cf. Table 1) by providing a short description followed by an overview of the relevant publications and a short conclusion. However, niches which are not addressed by a sufficient number of publications to justify an independent section are discussed in Section 3.3.9.

*3.3.1. Unit testing*

To apply unit testing, the system under test is separated into small pieces which concentrate on individual tasks and functions [S139]. Subsequently, test-cases are generated for each identified piece which stimulate the process and compare observed/expected behavior; during test executions the process is typically isolated from its partners by abstracting its interconnections.

The oldest discovered publications on unit testing are inspired by the software engineering domain. For example, [S1] proposes BPEL-Unit, which transforms a BPEL process into Java classes and methods to apply unit testing using JUnit – a Java unit testing framework. This approach provides great flexibility because the behavior of each process activity, gateway, and node (i.e., all process elements) can be described in source code and tested with source code based unit tests. However, it's a substantial amount of work to program a sufficiently detailed process model behavior simulation and we assume that this will not be applicable by typical modeling experts which most likely are not that well trained in programming. [S27] manually specifies tests that contain input and expected output data along with the expected execution paths for Microsoft's .NET Windows Workflow Foundation engine. Their approach can not be applied on other languages or engines which limits its applicability. However, they provide the possibility to 'simulate' the process execution based on the configured test-case input data. Hence, programming is only necessary if external partners should be simulated during test-case execution.

The following publications are exploiting that BPEL processes frequently communicate using SOAP messages that can be intercepted, modified, or injected using, for example, a proxy server. Hence, the proposed approached are able to fully control and simulate process executions and process partners by forging appropriate messages.

In [S30] a BPEL test framework is described. However, large parts of the framework are only proposed for future work and the test-case definition approach requires an extensive amount of manual preparation (e.g., manual definition of test data and expected results). Manual test-case definition, by describing the exchanged/expected messages, is also applied by [S41]. The publication is only concentrating on messages (i.e., order, content, amount) and is not specifying how the intercepted data can be efficiently analyzed and also does not observe the internal process execution operations. However, [S41] provides a neat way to define the abstracted simulated behavior of external process partners by specifying it as BPEL processes. So it can be expected that the modeling experts would be able to create those partner abstractions without requiring too much additional training. But, defining the simulated partner behavior – presumably – still takes a lot of time. [S40] concentrates not only on testing composite processes, using BPEL, but also on calculating properties such as test coverage (i.e., to assess which parts of the process are tested to which extend). Therefore, a 'Semantic Flow' model, which combines the data and control flow of composite web-services, is generated and analyzed. However, all test-cases must be defined manually, the paper does not contain any evaluation, and the Semantic Flow is generated based on a separate Web Ontology Language (OWL) based description of the process. Hence, two entities must be manually updated when modifying the process, the process model and its OWL based Semantic Flow representation. Likely both models will start to derivate over time. A similar concept is also in use by other authors (cf. [S32–S34]) to map BPELs block-structured process definitions onto graphs.

The previously discussed publications are defining their test-cases manually. Manual test-case definition can become a very expensive task, a challenge which was also recognized in [S42]. Hence, [S42] proposes to extract possible operations, supported by the BPEL process, from a Web Services Description Language (WSDL) file. However, the testers must still manually define appropriate test data and expected results. Hence, an in depth knowledge of the domain and expected behavior is required and multiple domains must be mastered to define, for each WSDL operation, expected response properties such as response times, HTTP codes, or XPATH rules. So test-case generation is still knowledge and time intensive.

However, other authors show that a more extensive automation is possible. For example, [S32] auto generates unit tests by using an extended control flow graph (XCFG). The graph is created based on the BPEL process specification to extract possible execution paths which form the individual test-cases. Subsequently, a constraint solver (BoNuS) is used to calculate input data (booleans/ integers) which induce the process to follow the test paths. However, the presented approach ignores the processes data flow and large parts of the BPEL standard as a 'workaround' for state explosion problems which occur during the test-case generation and this approach is, therefore, hardly applicable on real world process models. [S78] proposes a similar graph based test path genera-tion approach which applies McCabes's cyclomatic complexity [36] algorithm to check if all possible paths were found. However, the test data consists only of randomly generated values (including partners that are abstracted to return random data). Hence, test-cases can be generated, however, the expected outcome/behavior (i.e., expected execution path or results) must be specified manually and can most likely not be achieved based on the generated random input data. So this approach only identifies the possible execution paths but most of the hard work (e.g., which input data to use or which outcome to expect for specific input data) must still be done manually.

One alternative to defining tests by describing input values and expected behavior, is to specify process patterns and later check for their existence (e.g., specification/requirements) or non-existence (e.g., anti-patterns). A pattern based data flow testing approach is proposed by [S66]. They analyze the potentially exchanged messages and check for data flow anomalies by matching for a number of predefined patterns (e.g., are all message parts defined before sending it). However, test-cases which, for example, proof the problem or allow to verify if the problem was fixed can currently not be generated. Hence, all dynamic aspects (e.g., messages returned by partners) can not be integrated into the analysis. Patterns can also be applied onto the control flow. For example, [S70] defines patterns that are capable of checking if activity X follows after Y. Therefore, common process model control flow errors/weaknesses were identified and if the pattern matches then most likely a potential error was found. However, it suffers from the same drawbacks as [S66] approach (which, in general, also apply to all following none customizable pattern based approaches).

A contradictory approach is proposed in [S67] by defining *anti-patterns* (e.g., a specific XOR-AND element combination that leads to deadlocks). An anti-pattern should *not* be matched at a process model. They argue that their matching algorithm is faster than competing ones (e.g., BPMN-Q [37]). However, they transform the process model into a language independent representation and therefore lack the support for language specific features and struggle with complex process models because their language independent representation does not support all possibilities provided by, e.g., the whole BPEL standard. A comparable technique is proposed at [S68]. However, the identified static patterns must be matched manually which is time and knowledge intensive. Hence, their approach can hardly be applied on larger process model repositories and should be combined with existing matching algorithms.

The pattern focused publications mentioned above are only providing a fixed predefined set of patterns. However, to support individual requirements it's important that a tester can created custom patterns. Approaches which allow to define custom patterns are presented in [S43–S45]. In [S43,S45] the authors argue that current pattern description languages are too complex and therefore a new one was designed called PROPOLS (based on the Web Ontology Language). Their test algorithm transforms the patterns into Finite-State Machines (FSA). Subsequently, every possible execution order of the process is simulated and the states of the patterns FSAs is updated accordingly; it's required that all FSAs have reached their end state at least once to indicate that each pattern is supported. However, we assume that pattern definition and its parametrization is quite time intense because [S43,S45] require to model all states, relations, and the expected behavior of the internal process and and its partners (i.e., creating Java source code as process partner simulations) by hand. Also the authors are *not* providing any user study based evaluation to prove that their approach can more easily be applied than existing ones – a challenge which should be addressed in future work to assess the validity of their assumptions.

[S44] uses a subset of the UML 2.0 specification to describe patterns (to check for specific properties) which are automatically transformed to state machines. Process executions are monitored and state changes are triggered according to the captured events (i.e., which messages are currently exchanged, hence, none static runtime analysis is supported). A property is marked as fulfilled if its state machine reaches its end state; or marked as failed if it reaches a predefined error state. Yet another language (Prolog) is used to define control flow patterns in [S122] to enforce specific modeling rules. However, [S122] only supports control flow pattern and static process model analysis. Interestingly enough they not only use their approach to identify potential errors but also to enforce modeling experts to follow, for example, organization wide modeling guidelines.

We found that the majority of the approaches require that the modeling experts learn novel pattern description languages, hence, an extensive amount of training is necessary. The lacking support for cross-organization processes enforces that all processes executions and monitoring must be executed on the same server. However, this is unlikely the case for distributed process models were each integrated process partner can call numerous supportive services. Again, automatic test-case generation is not supported yet and only mentioned as future work by the pattern based unit testing approaches which results in high test-case generation costs.

Unit testing is an important testing technique which provides a rapid approach to define individual tests. Early approaches, such as [S1], were clearly inspired by the software engineering domain, whereas later publications became more independent in terms of the used techniques and approaches. However, the presented approaches struggle with separating the process models into individual and separately testable units. For example, unit tests at the source code level are applied on the 'smallest testable part of an application'[38, p. 244] (i.e., on method, interface, or class level) while for process models the 'smallest parts' are typically whole execution paths which integrate a diverse set of complex activities. Hence, generating appropriate test-cases/data and interpreting test results becomes harder because of this increased complexity. One additional major challenge remains: the high cost to create all the individual test-cases, because a majority of the described approaches rely on time consuming manual test-case definition. In order to minimize that effort it becomes necessary to auto generate tests and the according test data.

*3.3.2. Test-case generation*

The ISO/IEC/IEEE 24765:2010 standard, defines a test-case generator as '*a software tool that accepts as input source code, test criteria, specifications,* *or data structure definitions; uses these inputs to generate test input data; and, sometimes, determines expected results*' [39, p. 29]. This applies, in general, also to the process domain. However, the test algorithms utilize process models instead of *input source code*. This section starts by discussing how execution paths, a foundation for most test approaches, can be extracted from process models. Subsequently, those execution paths are reused by data generation algorithms (cf. Section 3.3.3) to determine appropriate test input data.

[S33] proposes an approach to extract possible execution paths (and therefore possible test-cases) from a BPEL process model. Therefore, the model is transformed into a BPEL Flow Graph (BFG). A BFG is a Control Flow Graph (CFG) which is extended to support BPEL characteristics, however, large parts of the BPEL standard, such as exception paths, are still missing. The BFG is analyzed with graph search algorithms to extract possible execution paths. Note, that the paper is missing a comparative evaluation and is mostly just 'discussing' how potential state explosion problems could affect the presented techniques. However, a solution for this challenge is not presented. [S69] applies a similar technique called PCFG. The additional P represents the integration of connected partner process models – their private process view must be accessible, which is rather unlikely at cross-organizational processes. However, they are also proposing rare ways to measure test quality by analyzing test-coverage (i.e., for example which process activities are covered by the existing tests) and support their work using one of the few identified case-studies. Execution paths are extracted by applying data flow centered extraction criteria (e.g., paths that cover variables which are forwarded to partner processes). Both approaches [S33,S69] are extracting all possible execution paths which can lead to a state explosion problem [40] – more tests are generated than the system can check in a reasonable amount of time.

Hence, [S113] and [S111] present a solution which is based on extracting only the most important paths, called 'business rules'. However, a user must guide the path extraction algorithm (to identify the important paths) and has to manually specify test data. Additionally the support for the extracted rules must also be checked manually. Hence, its unlikely that the proposed approach can be applied on large real world process repositories, especially, because all work is done manually and all concepts to automate this approach are 'postponed' as future work by the authors.

The previous three approaches utilize a *custom* graph representation. Other researchers, however, are converting the yet to be tested process models into existing modeling language such as UML. For example, [S46] transforms BPEL models into UML 2.0 activity models. The test paths and therefore potential test-cases are extracted from the UML models, based on various criteria (e.g. full path coverage), by applying a depth first search. However, the publication is missing detailed information how the test-cases are executed/compared with the process model to identify derivations and the utilized activity model based behavior representation requires a substantial amount of manual annotations and domain knowledge to create the required formal specification of the expected behavior. [S100] proposes a similar approach by generating an UML state dia-gram from the BPEL process and an extended FSA from the associated WSDL file. Both representations are combined to an EFSM-SeTM model which indicates how different partners/activities are related (based on the BPEL/WSDL file) and which variable format is in use (based on the WSDL file). The possible execution paths are automatically extracted. However, it's not specified how derivations from the expected behavior (i.e., faults) can be identified and only the test-paths are generated automatically. Hence, a substantial amount of work is required to generate test data, select the yet to be tested services and appropriate test-cases and a majority of the components which are required to create a complete test framework are postponed by the authors as future work. [S48] transforms process models defined in custom domain-specific languages into UML models to manually add UML U2TP stereotypes – U2TP is an UML extension which allows to annotate details that are helpful to conduct tests. FOKUS!MBT then generates tests based on this extended UML models. This approach benefits from extending U2TP because it's already standardized and documented and therefore it's likely already known by testers. However, modeling experts most likely will not have used it yet because the interest in the industry is relatively low, as most UML standards (and the presented extensions) it's quite complex, and the authors identified some inconsistencies when comparing it with the latest UML releases.

Other authors also transform UML models. In [S47] a UML model is transformed into a Color Petri Net (CPN) representation to apply a random walk algorithm (randomly choosing the next valid step) until all possible execution paths are extracted. The paper starts by transforming UML Activity Diagrams, however, most likely the presented test path extraction approaches can more frequently be applied directly on Petri Nets because those are more common at the process modeling domain then Activity Diagrams. Additionally, the authors did not define ways to identify test-data which would be required to create complete test-cases from the identified test-paths. However, the authors address, for example, concurrency and proof that they are able to support concurrent execution testing by finding each feasible activity execution order.

Generally, a transformation to Petri Nets is popular because Petri Nets provide a strict and formal representation that can be used as a *workaround* for the nondeterministic behavior of modeling languages such as BPMN [41]. [S71] transforms a BPMN modeled into an Algebraic Petri Net. Tests are randomly generated based on manually specified rules, defined in Hennessy-Milner Logic (e.g., a test must cover activity X). The combination of a rough Petri Net and fine granular rules allows to automatically generate focused tests which cover specific parts or situations were the testers are most interested in. However, only the test paths are extracted and most importantly the paper is missing an evaluations section and it is doubtful if a typical modeling expert would be able to write the semantic rules out of the box. [S72] proposes to test BPEL processes by using a CPN which represents the required properties. CPN models support data, hence, a WSDL file is analyzed to include the type/format of each variable. The execution paths are automatically generated based on coverage requirements and a manual activity prioritization. The authors promise that their approach is able to reflect data and data variable constraints better than existing approaches because the requirements and constraints for each data field can be manually defined. However, the required CPN models must also be manually created or (when assuming that it's already 'existing' because the designers used it for design verification) at least extended with custom rules to model data requirements - which increases test costs. So, it would be beneficial to utilize auto-generated knowledge sources such as execution logs to automatically extract and generate test data, rules, and constraints based on the recorded 'normal' behavior to ensure that the process behaves consistently after, for example, a process partner was changed.

An automatic transformation from BPEL to CPN is also used at [S74]. The authors are concentrating on concurrency and are showing the efficiency (i.e., how state explosions are prevented) of their approach in concurrent situations using multiple examples. The execution paths are directly extracted from the CPN by graph search algorithms. BPEL is missing a graphical representation. Hence, [S52] transforms BPEL into BPMN to create a visualization and to extract execution paths. However, the presented test path extraction is not particularly innovative (i.e., already presented in previous work).

Above transformation approaches typically transform one process modeling language into another one to gain some advantages (e.g., to reuse existing algorithms). However, another alternative is to transform the model in an abstract representation such as timed state machines. A timed state machine is used in [S26] to model details (e.g., timeouts) into the test generation approach. Test paths are then extracted by applying the Hit or Jump algorithm. Although, the main contribution of the paper is concentrating on model transformation than on process testing, it still leaves many open areas of improvement to complete the formal representation of the whole BPEL standard.

[S109] even modifies the process model to reduce the amount of necessary test-cases. They are applying the Category Partition Method (CPM) which requires that the process model gets substantially annotated, for example, by dividing each input variable into multiple partitions. Each partition represents value ranges of equivalent data which can be used for test data generation. Additional metadata (e.g., priority, required combinations, or logic rules) are assigned to each partition and to activities. All this information controls the automatic test path extraction and, therefore, the test-case generation. The tests itself are generated almost automatically, however, the preparatory work and the required domain knowledge is enormous and we assume that it should be possible to automatically extract a noticeable amount of the information from various knowledge sources (e.g., how often each patch is executed can be taken from recorded execution logs). In addition, the original process model is extended and modified to integrate information which is required to generate the test-cases. However, this mixes up multiple concerns into the process model and therefore hardens its maintenance and development. [S35] proposes an approach which is, similar to CPM, also inspired by software testing approaches. Hence, the BPEL process is transformed into a Symbolic Transition System (STS) to generate a Symbolic Execution Tree (SET) that contains all

possible execution paths. The main difference between a normal and a symbolic execution is that no concrete values are used, instead variables are replaced with symbolic formulas. This allows to minimize the amount of generated tests because each symbol can represent a range of possible values. Hence, this approach deals with complex XML based data variables without suffering from state explosions. However, the presented work requires a careful transformation of the process models into a processable formal representation and again only a small subset of the whole BPEL standard is currently addressed by the authors which limits the applicability of the this approach. Section 3.3.3 discusses how paths extracted from the SET can be populated with real values to generate executable tests.

A major disadvantage of transformation based approaches is that the tests are no longer generated on the 'real' process model, but on an abstracted representation. A diversion between the real model and the tested model is possible, an effect which can be prevented by generating tests based on the original process model. Additionally the transformation based approaches frequently only support a subset of the transformed modeling standards and can therefore, presumably, not be used on complex real world process models which most likely rely on some of the unsupported modeling language aspects which can lead to a complete failure of the test generation approach or incorrect test results.

Hence, [S78] proposes a path generation approach which extracts possible execution paths directly from a BPEL model. This can create long execution paths, especially at loops, so that [S54] proposes an algorithm that respects a configurable maximum path length. However, the papers only contain a motivational example and are missing an evaluation to proof the feasibility of the presented approaches. Overall, the papers only extract test paths and must be combined with third party tools to generate, for example, appropriate test data to create complete test-cases.

[S79] applies scatter search. The generic search algorithm was extended to address BPEL characteristics such as fault handlers. Hence, a sub goal (e.g., a transition) is selected and the search algorithms searches for paths which cover this goal. The authors state that they are able to identify 100% of all possible execution paths. However, they only support a fraction of the whole BPEL standard and only compare it to a very limited testing method (random testing) during the evaluation. Hence, the presented approach is 'unsurprisingly' better than the compared one and the significance of the evaluation is limited. Note, that we found similar evaluation quality issues at multiple process model testing papers.

[S132] proposes to utilize message sequence graphs which are generated from BPEL models. All possible sequences are extracted, populated with random data and executed while checking for exceptions. However, the presented approach does not specify any sort of expected behavior which could be compared with the observed one. Hence, they are only checking for generic errors such as thrown exceptions and can therefore be applied on systems with unknown behavior and documentation. However, this also can quickly result in false positives because, for example, there is no way to describe expected exceptions.

Another concept is to exploit the counter example generation capabilities of verification tools (cf. [S101,S102]). Therefore, the process model is transformed into a language (e.g., PROMELA) which can be processed by verification software like SPIN. The tester then specifies *negated* properties, for example, in Linear Temporal Logic (LTL) [42] (e.g., that transition X can *never* be reached). The verification software will then use this property (e.g., Quality of Service requirements) and generates a counter example (how transition X in fact can be reached) that can be used as a test-case. This approach greatly benefits from a wide range of optimizations and research which were already invested into existing formal verification techniques, for example, to deal with state explosions. However, this promising foundation requires the modeling experts to learn new none modeling related languages, for example, to describe test properties. Again, the authors are only able to prove the correctness of their model transformations approach for a very limited subset of the BPEL standard.

The previously described approaches (except for [S69]) concentrated mainly on generating execution paths based on the control flow (the order in which the individual activities are executed). However, a process model also contains a data flow that describes how data is exchanged between different activities.

Hence, test paths can also be extracted based on the data flow using adapted coverage criteria such as all-du (the define/use paths of all variables). An all-du path extraction criteria is applied in [S107,S115] and evaluated in [S116] for BPEL (including solutions for BPEL characteristics such as death-path elimination [43]). BPEL does not contain an explicitly modeled data flow, hence both papers first 'detect' probable data flow paths. However, both papers are missing a thorough evaluation and only detect 'basic' paths (e.g., ignoring fault handling, at least concurrent model aspect are supported by [S115]). [S108] widens the covered area by considering the XPath rules which are used at BPEL to, for example, extract variables from XML messages. Therefore, the BPEL process is converted into a CFG that is extended with XPath Rewriting Graphs (XRG – a graph representation of a XPath rule) to create a X-WSBEPL model which is used to extracted data flow paths. XPaths are an important data handling artifact in most BPEL processes which were ignored by previous work. Hence, this approach is one major step forward to fully test all aspects of BPEL processes, the authors even performed one of the few identified case study to evaluate their ideas. However, the data dependencies of the individual activities are ignored by the authors. It would be promising to combine, for example, the approaches presented in [S107,S115] with this technique.

[S12], instead, calculates a data flow graph from a BPEL model and then extracts all possible paths. However, the authors only utilized the technique to measure change impact and to identify test-cases which cover changed execution paths and process areas. Hence, it would be interesting to refine the approach to see if can also be applied successfully during test-case generation, for example, to generate tests specifically for changed process parts.

Until now, concurrency (multiple branches can be executed in parallel during a process instance execution) was not explicitly discussed – although implicitly supported by many process model testing approaches. Concurrency, is a challenging topic because it leads to a great variability in the possible activity execution order – increasing the number of necessary test-cases and the risk of a state explosion. ParTes [S123] was specifically designed to counter that risk. ParTes is limiting the amount of generated execution paths by checking if the executed activities (of e.g., two different parallel branches) are communicating with the same or different sub systems. Only if they are communicating with the same sub system (i.e., process partner), then a race condition can occur and all possible execution paths are generated. The process model is adapted/tested by explicitly modeling the various execution paths (XOR gateways allow to individually select each path). Hence, this approach reduces the risk to be affected by state explosions. The authors are currently only applying this on concurrent activities which communicate with external services. However by integrating data paths into the approach (see [S107,S115]) it should also be applicable in a broader more general way when applied on all concurrent control flow execution paths.

This section provides an extensive overview about the existing test path extraction approaches. Two major aspects that can be take from this section, that are a) concurrency and connected state explosion problems are already acknowledged by existing work and b) that formal test path generation approaches exist but all of them struggle with proving the feasibility of their approach on the whole respective process modeling standard (e.g., BPEL) they are aiming on. The presented approaches are important for the following sections because they all require test-cases and are therefore are frequently reusing the presented concepts. However, each test-case also requires test data, a topic which is addressed in the following.

*3.3.3. Test data generation*

Test data is the second most important component required to 'assemble a test-case' (compare, Section 3.3.2).

A primitive approach to generate test data is to do it *manually*. Hence, a user has to study the objectives of each test-case and select appropriate input data (cf. [S27,S28,S30,S31,S40,S46,S126]). However, this is substantially increasing the cost to generate tests.

Hence, an automatic data generation is required. A simple approach, used by [S71,S78,S79,S99,S109,S127,S128,S132,S135,S136,S138,S139,S142,S159], is to generate *random* test data or complete random messages for robustness tests (cf. [S127,S128]). Random generation can easily be applied on numbers/booleans but becomes more complex for strings which frequently must follow a specific structure. Thus, strings are frequently only randomly selected out of a predefined list (cf. [S135]).

Despite numerous efforts (specifying value ranges [S78] or defining partitions [S109,S138]), it's still 'random' data and so it can take a while to generate a value combination which triggers all guards correctly so that the desired test execution path is followed. Of course, it's possible to adapt and modify the yet to be tested process so that each test path can be selected individually (cf. [S123,S141]). However, this might distorts the results and a more focused approach is needed. Hence, [S105] proposes to start with a manually defined value which can later be optimized (so that it fulfills the guards/conditions of the test path) by using a genetic algorithm. A reversed approach is used in [S106] to identify value combinations which would violate Service-Level-Agreements (SLA) and to generate tests which prove that the SLA is not fulfilled.

Search based approaches can identify the same 'optimal' input data for multiple test-cases. Hence, [S104] utilizes *tabu search* (generated values are noted on a tabu list and ignored for a certain amount of generations to increase the likelihood that a new value is chosen). In [S80] a scatter search algorithm is applied that combines evolutionary ideas (genetic algorithms) with tabu search.

An alternative to search algorithms was found in constraint solvers. Constraint solvers (used by [S32,S33,S35,S36,S48–S51,S54,S74,S77,S140,S145]) are based on algorithms which get as input some constraints and output a value which fulfills, if possible, all of them. Hence, each extracted execution path is searched for guards/conditions which must be fulfilled so that the process instance follows this test path. The identified conditions are utilized by a constraint solver to generate appropriate test data. Similar results can be achieved using verification software and their counter example (cf. Section 3.3.2) generation capabilities [S101,S102] – typically, path and data are generated at once.

This and the previous Section 3.3.2 provide the requisite tools to automatically create test-cases. Those will be used not only at the initially described unit testing approaches but also at the testing techniques addressed in the following. We see search algorithms and constraint solvers as the most valuable approaches for test data generation. Both approaches can generate artificial test data by utilizing the constraints and rules described at process model gateways as a foundation for test data generation. However, frequently this base set of rules must be manually extended, for example, by annotation pre- and postconditions at each process partner or even activity which is time intensive. Hence, it would be promising (e.g., for regressing testing) to generate the required extended ruleset – which guides the data generation process –, for example, from recorded previous process model instance executions (e.g., using recorded execution logs) to identify, for example, typical data ranges and data structures which can be used by the presented data generation approaches to create realistic test data.

*3.3.4. Integration testing*

Integration testing is defined as '*testing of programs or modules in order to ensure their proper functioning in the complete system*' [44, p. 8]. Hence, integration testing can check if a process can correctly interact with their partner processes and services.

Therefore, [S126] proposes to integrate sensors into all process instances (including the partners) to observe test executions. Subsequently, the expected behavior is described in High Level Petri Nets to extract execution paths; test data is added manually. The tests are executed while checking if the observed behavior mimics the expected one. However, it is unlikely, in real world scenarios, that an organization is allowed to integrate sensors and new logic into important processes of partner organizations and the approach was also not evaluated.

[S105] annotates the process model with meta information (pre/post-conditions at activities and constraints for the exchanged values). Subsequently, the whole process is converted into Java source code to apply Java Path Finder to extract execution paths. The test data is generated with genetic algorithms and model checkers (counter example generation). Finally, each test is analyzed and the exchanged messages are extracted and used as integration tests by comparing the returned values with manually specified constraints. The main idea is to identify execution paths with as many partner interactions as possible because those most likely lead to faults. However, the potential state explosions during the search are, based on the paper's authors, negated by applying genetic algorithms and formal approaches. However, the approach is computational intense, requires a substantial amount of manually annotations and domain knowledge – which increases testing costs, and does not provide a through evaluation (i.e., just some small examples).

[S36] proposes to use UML models (e.g., activity model and interface model) as contracts that restrict the behavior of the process and its partners. Therefore, each activity gets manually annotated with pre- and post-conditions. The test-cases are automatically generated (path extraction, guard/condition based data generation, cf. Section 3.3.3) and executed while comparing the observed behavior with the conditions. The proposed approach is promising. However, large parts of the whole concept are just described as potentials solutions and postponed as future work, for example, the approach is missing a way to compare the observed behavior with the expected one to detect derivations. [S37] converts the process into a state diagram to extract the hierarchical order in which the partners are called at specific execution paths. The extracted call order is used to manually specify test messages and expected return values. The only automatically executed aspect is the test path extraction, hence, even the test execution and the comparison of the results with the expected values must be done manually. We assume that it is not applicable on complex business process models because those manual time intensive steps lead to high costs.

Potential partner processes are reviewed in [S38] to identify if they can be integrated into the already existing internal process. Therefore a Orchestration Participant Testing strategy (OPT) is generated that consists of a graph representation of the process and manually defined properties (LTL statements) which allow to create tests based on model checkers (counter example generation). The further steps are comparable to [S105]. The presented approach is work intensive. Why? Because each test is 'manually' generated, because the presented automatic test generation always requires a complex manually specified test property. The possible test properties are limited and likely not suited for real life process models and the model checkers only supports a subset of the BPEL standard. Hence, the test generation approach is promising but must be extended and evaluated in regards of usability and model language support.

[S39] addresses the same challenge by identifying execution paths which contain interactions with the potential partner. Finally, the test data is defined

manually and the potential new partners are bound. Then the tests are executed and simple error conditions are checked (e.g., if error messages/exceptions are observed). Finally, the results should allow to assess if the new partner can be correctly integrated into existing potential bindings (i.e., each service integration point can be replaced by multiple potential concrete candidate services at runtime). Data flow metrics are utilized to reduce the amount of potential bindings which must be evaluated by checking if and how each service is connected with each other services. The whole testing process is widely automated and the authors provide a comprehensive evaluation.

The publications presented in this section mainly concentrated on ensuring if partners can correctly be integrated with the process under test by applying integration tests. Section 3.3.5 shows that a similar objective can also be followed for non-functional requirements. We found that the support for cross-organization scenarios is quite high at the identified integration testing approaches. Hence, it would be interesting to explore existing integration testing concepts and see if their assumptions, ideas, and concepts could also be applied on other testing domains such as unit tests to strengthen the support for cross-organizational scenarios in such test domains.

*3.3.5. Quality of Service testing*

Quality of Service (QoS) is defined as a '*set of quality requirements on the collective behavior of one or more objects*' [45, p. 2]. Hence, the following publications mainly concentrate on ensuring that non-functional requirements (e.g., availability or response time) are fulfilled. In this context, cross-organizational scenarios pose a particular challenge because external partners can change quickly and, initially, unnoticed.

[S57] and [S56] are addressing this challenge. Hence, [S57] proposes to annotate each process activity with its expected completion times (one for each potential partner implementing this activity). Genetic algorithms then search for input data and partner combinations which would violate a predefined maximum completion time. The presented approach requires that the 'real' response times of each service are available and it is assumed that those are completely equal for each request (i.e., independently from the request data which is, likely, a unrealistic assumption). Hence, currently a tester has to manually defines the response times in a coarse grained manner (which is error prone, time intensive and requires a in depth knowledge of the process partner services), but the same information could also be extracted by analyzing historic execution logs or by sending randomly generated test messages to the services. Hereby, it is also possible to create a more fine granular configuration and more realistic results for average and worst case scenarios which increases test quality.

A similar objective is addressed in [S56] by applying pair wise testing to reduce test redundancy by 99%. Pair wise testing is a combinatorial testing approach that creates unique pairs of input data (i.e., partners which can be chosen from to implement each activity) and then tests those combinations [46]. Therefore, a feature diagram is created that contains the interactions and features (e.g., delay or cost) of each partner. Then the process is converted into a concurrent programming language (ORC) to generate tests which indicates a possible violation of a guaranteed SLA. However, concurrent process model aspects were only discussed theoretically and, therefore, the results fall short of what is possible. The authors also conducted two case studies. However, they only compared their approach with a random testing approach while a comparison with more advanced approaches, e.g. [S57], would provide better insights.

A less complex, but still powerful, approach is used in [S21] to check if a process model respects a maximum completion time in each possible execution path (classic verification properties such as deadlock checking are also supported). Counter example generation techniques are applied to generate the possible paths while timed state transition systems describe the approximated execution times of each activity. The approach requires that the BPEL model is extend with custom annotations. However, the BPEL model should, at least based on our assumptions, mainly contain data which is relevant for process model execution. So, it would be interesting to check if the relevant information can be stored in a separate place, e.g., the Web Service Timed State Transition System model (WSTSTS). The WSTSTS is core component of this approach and is generated by transforming the BPEL process. Hence, if there is already a separate model required then it could be also used to separately hold the test data.

The previous approaches did not consider that SLAs can describe complex dynamic behavior (e.g., a 1 MB file is processed within 2 minutes but a 3 MB file needs at most 10 minutes). Hence, [S106] focuses also on variables and input values. Therefore, each partner must create a fine granular SLA that specifics value ranges and their guaranteed execution times. A genetic algorithm then searches for potential SLA violations. The presented approach quickly identifies problematic areas. However, we assume that the required very fine granular SLAs are not available in real scenarios. This challenge also applies to the following approach. [S25] proposes to evaluate QoS specifications based on autogenerated stubs which reflect the SLAs of the integrated partners (e.g., maximum calls per minute or response time). The stubs are replacing the real partners and a typical workload is simulated to check if any of the partner SLAs would be violated or if the overall performance of the process does no meet its own SLA.

The previous approaches are based on the expectation that each partner process provides detailed SLAs. However, the papers are not specifying (except for [S25] which uses WS-Agreement [47]) how SLAs could be defined in a standardized machine readable format. Hence, the approaches are hardly applicable on real world scenarios and upcoming researchers should consider to reuse existing SLA specification languages or propose new ones (or extensions to existing ones) which allow to, e.g., conduct case and users studies to compare the efficiency of the various approaches.

A further attempt is based on monitoring process instances. Two directions can be observed: a) a fingerprint of the typical behavior is generated and compared with subsequent executions or b) running instances are monitored and compared with manually defined rules. [S22] uses direction a), and generates a fingerprint based on support vector machines. The presented approach can currently only be applied on hand piked variables which must only contain numeric data. Hence, the approach is not yet suitable for large scale monitoring which would be required in real world scenarios. Direction b) is utilized in [S23,S24]. Hence, [S23] modifies the process model and adds sensors after and before each activity. This allows to observe/check execution times and variables. For each sensor a manager was designed which compares the observed data with predefined rules (i.e., test-cases). However, all the rules must be defined manually in a custom language which, because of the huge amount of sensors, ends up at a substantial amount of preparatory work. The authors are only providing a performance evaluation because they have noticed that adding a big amount of sensors can also generate a large amount of traffic. Therefore, [S24] concentrates on finding the optimal places to integrate the smallest possible amount of sensors into a process model to fulfill predefined coverage criteria. The sensors are configured with manually defined monitoring rules and the activity which most likely violated a rule is reported. However, [S23,S24] almost exclusively concentrate on execution times. Hence, it would be promising to combine them with approaches presented in, for example, [S22], to cover a broader area of risk/data based quality of service rules in combination with execution time based quality of service rules.

The dynamic nature of processes is addressed in [S10] by proposing to define constraints (e.g., in activity structure or temporal) as some sort of SLA which enforce the modeler to support basic properties. Hence, a normal user can start adapting a base process model so that the process model can be dynamically adapted while specific 'guaranteed' properties are still ensured. Currently the constraints must be generated completely manually and this approach is therefor time intensive. Hence, it would be promising to define an automatic two sided transformation between SLAs and checkable constraint rules (and vice versa) which could be applied in such an scenario to reduce constraint generation workload and to ensure that the public SLAs are always up to date.

This section concentrates on generic quality of service aspects like availability or data quality. We found, that the identified publications are frequently not exploiting existing knowledge sources such as existing service level agreements so that all rules and conditions are generated newly. Hence, we assume that the separately stored rules/conditions will likely derivate from the public documentation over time (which, e.g., confuses external partners) and also an unnecessary amount of work can be saved by generating an automatic transformation between SLAs or execution logs and the enforced quality of service rules. The following two approaches are connected to quality of service testing, but form specialized groups that concentrate specifically on performance testing as well as robustness testing.

*Performance testing.* Performance testing tries '*to evaluate the behavior of a test item under anticipated conditions of varying load*' [48, p. 8]. Hence, it analyzes performance indicators (e.g., response time) or checks if the system is in a consistent state even under high load.

[S65] proposes to simulate process execution while measuring the response times of all partner systems. Overall, the recorded information allows do evaluate the performance of entire business process environments. However, all simulated executions and test message must be defined manually, which is time intensive. Hence, to apply this approach in real world scenarios, it is necessary to increase test generation speed, for example, by automatically extracting correct message orders from business process models. If performance testing should be done on an already finished process model (e.g., during maintenance) then the message could also be automatically generated from recorded previous executions.

The concept of aspect oriented programming inspired [S64] to propose an approach that weaves timers (e.g., before/after each activity) into XML files which store process models. Subsequently, those timers are used to measure execution times. However, this only generates the necessary data for a manual posteriori analysis because the authors did not specify any sort of rule language or concept to define constraints/tests which evaluate the generated data - which greatly limits the applicability of this approach. One of the main advantages of this approach is that the timers are weaved dynamically into the process model just before the execution so that the model is *not* cluttered by none business related elements. Note, that this still can lead to issues during the execution (which are currently not addressed), e.g., how to handle situations where the sensor manager, that is collecting all data, is not available. For example, this could 'block' or slow down the process execution because a sensor could, e.g., wait until the connection to the manager is closed because of a timeout.

[S142,S159] also asses (in additional to e.g., the response time) if the observed behavior still respects a predefined specification (modeled as a timed automaton) if the process is under load. A configurable amount of processes is created that are executed using random input data. All partners are replaced with stubs and the incoming requests are logged and compared with a specification (e.g., to check the message order).

Note, that random data might not be perfect for this solution because it maybe substantially differs from real life data. Hence, a search or constraint based approach, cf. Section 3.3.3, might be beneficial. Alternatively, recorded old execution logs could be replayed in masses. This would be especially come in handy because they represent real world use cases and should be available at certain scenarios, e.g., when a former process partner is replaced with a new one.

The presented performance testing approaches allow to evaluate if the system behaves as expected at various load levels. Here [S64] proposes a interesting aspect oriented programming, cf. [49], based approach that is able to integrate process model extensions, that are required for testing, dynamically when preparing the model for execution. Hence, the testing aspects are not mixed with the business aspects and so the model and its tests become easier to maintain. It would be promising for future work to check if this concept can also be applied on other testing approaches. However, performance tests can also be seen as some sort of robustness tests which ensure that a process is consistent even at high load. Nevertheless, specialized robustness testing concepts are discussed in the following.

*Robustness testing.* Robustness testing assesses if a process behaves correctly if it has to deal with mutated input data or flawed environmental conditions such as databases that report errors instead of the expected information. In computer science, robustness is defined as '*the degree to which a system or component can function correctly in the presence of invalid inputs or stressful environmental conditions*' [39, p. 313].

We found that this topic is approached from two angles: a) publications which describe mutation algorithms that can create corrupted input data or flawed environmental conditions and b) publications which use those mutation algorithms and discuss how to conduct robustness test on a process model.

[S129] proposes to automatically adapt the exchanged messages and database connections (e.g., by applying data corruption or adding delay). The processes are evaluated by measuring how many activities are executed (lower = better) before the process instance recognizes and reacts on the mutations. This approach basically assumes that each type if incorrect should lead to a 'termination' of the process. Hence, it can lead to false positives when dealing with a very robust process which, for example, is able to repair ingoing incorrect data or if the process is able to still continue its execution but then follows a specific error handling path to deal with the incorrect data. [S127] tunnels all messages though a proxy to, for example, delay or mutate them. They evaluate a process model by comparing if the process reaches the same end state with and without the injection of mutated messages, which would be the case in an *optimal* situation because then the process would not be affected by the mutated data. A comparable approach is proposed in [S49,S131]. However, the proposed three approaches also generate invalid XML data or messages which do not conform to the specified message interfaces and that should, therefore, be automatically rejected by the workflow engine. Hence this partly reduces the efficiency of this testing approaches in process oriented scenarios compared to, e.g., pure service oriented scenarios were the service developer must specify how invalid XML data should be handled. Overall, [S49,S131] provide the most complete solution (compared to [S127]) because they also allow to simulate classic network errors such as lost messages or services which are temporarily not available.

[S133] assumes that a process should always continue its execution, for example, by repairing corrupted messages. Hence, all process partners are abstracted and valid, invalid, expected, and unexpected exceptions and messages are automatically generated and sent to the process while observing its behavior. A major disadvantage of this approach and its fundamental assumption is that it is not checked if the behavior of the tested process model instance is still correct after, for example, a corrupt message got repaired or if the repairing logic was too optimistic and the process is now executing steps which were not foreseen by the message sender.

[S133] is a contrasting approach to [S129]. So, similar disadvantages exists

and a combination of both approaches would be promising to major the development in this area and to evaluate the synergetic effects of such an advanced solution.

In contrast, in [S134], the test engineer has to manually define each yet to be injected fault (e.g., incorrect parameters) and the process model instance state that should be reached after the injection. Of course this leads to an enormous amount of manual preparation and effort for each test-case so that this approach can only be recommended as an addition, because of its flexibility and fine grained configurability, to the previously discussed approaches.

A different approach is presented in [S130]. It's based on a symbolic specification (STS) of the expected behavior. The STS is transformed into an SET which is split to generate sub trees that are only addressing one specific BPEL operation. Later on, random input data is generated to call those operations and the behavior is compared with the STS and with custom rules (e.g., is the maximum response time below X). Note, that the author also utilized the same concept to perform Quality of Service testing. Hence similar drawbacks apply at both areas, for example, that the user would end up with two separate models (e.g., a BPEL and a STS based representation of the specification), which must be kept in sync and which require a diverse set of skills to be modified and developed. Hence, both models likely divert or contain different errors and it would therefore be advisable to assign the development of both models to independent experts which would also allow to identify vague specifications.

The described robustness testing approaches allow to assess the stability and robustness of the tested process models, and therefore contribute to the overall quality of service provided by a process model. We found that robustness testing frequently concentrates on a rather unspecific way of testing (e.g., messages or data are mutated randomly).

*3.3.6. Conformance testing*

The ISO/IEC 13233 standard defines conformance as the '*fulfillment by a product, process or service of all relevant specified conformance requirements*' [50, p. 29] while in the ISO/IEC 10641 documents conformance testing is described as a '*test to evaluate the adherence or non-adherence of a candidate implementation to a standard*' [51, p. 4]. Hence, conformance testing allows to determine if a process complies to some specification/constraints.

Conformance testing can be applied, for example, when a process model is adopted by a user on the fly during its execution while specific elementary requirements must still be met. [S91] tackles this challenges by providing predefined activity constraint templates (e.g., X must be followed by Y). Hence, a modeling expert defines some conformance rules which are enforced when the user applies modifications. However, this approach is only partly connected to testing and does mainly concentrate on enforcing specific constraints when process models are modified by hardly trained users (i.e., dynamic modeling). It would be promising to extend this approach, for example, by transforming specifications into constraint templates or by analyzing multiple revisions of the same process model and to automatically extract stable areas as constraints because those could build a important foundation also for future process model revisions. Finally, while the authors state that they wanted to propose a human centric approach they did not yet conduct a user study to prove the applicability of their approach in real world scenarios.

However, constraints can also be used to test processes models. Hence, [S28,S90] propose to specify constraints as a Timed Extended Finite-State Machine (TEFSM). Subsequently, existing test-cases are executed and the observed behavior is compared with the TEFSM. This approaches allow to integrate *time* into the tests. Hence a more diverse set of constraints is supported which can take expected or recorded runtime behavior into account, e.g., by checking if a response is sent after at most 30 seconds.

Another, comparable, automation based approach is proposed in [S44] and [S4]. However, defining an automation is complex and requires skills that, we assume, are not available at a typical modeling expert. Hence, [S44] proposes that the constraints are described using UML sequence diagrams. Later, those UML models are automatically transformed to an automaton. One of the main advantages of this approach clearly is that a subset of the frequently applied UML modeling languages is adapted to specify constraints. Hence, the expected users (i.e., modeling experts) are most likely able to apply the presented approach.

[S143] utilizes Petri Net models to describe the utilized specification, which also ensures that the required modeling skills are most likely available. The specification is transformed into a finite state machine and conformance is tested by generating checking sequences and looking for violations (e.g., states that should be reached but are not activated). In particular, a fast algorithm to perform those checks is presented by the authors. However, their approach is highly specialized and can only be applied on a specific type of petri nets (called 1-safe nets, limiting the applicability of this approach in real world scenarios where other modeling languages are frequently used) and the authors did not provide any kind of evaluation.

A tool is presented in [S99] that conducts conformance tests by generating random test-cases with random data based on a specification which was transformed into a TEFSM. The behavior of the process is observed and compared to the specification. We assume that the utility of this approach is limited. Why? First of all it only provides a inefficient test-case generation approach which is purely random. But most importantly it generates the specification out of the process model which is later tested (e.g., the timeouts used for the test-cases are taken from the timeouts specified at the BPEL model). Hence, the generated test-cases are testing one by one the behavior that is already specified at the process model and we assume that diversions from the 'specification' would only be observable if the workflow engine itself contains a bug (e.g., all partners are simulated too, hence, not even errors at process partners could be detected).

[S81], proposes a similar approach, with similar drawbacks, by converting a specification (Web Service Choreography Interface and Web Service Choreography Description Language models (WSCDL)) into a timed automaton. The UPPAAL model checker is used to check if the observed behavior of the process model follows the specification (checking only timeouts and message order) dur-

ing the simulation of random executions. This approach is one of the few which places more emphasis on integrating partner processes. However, each partner, and also the process under test, must be manually 'converted' into timed automata so that this approach is time intensive, expensive, and can hardly be applied during the maintenance of legacy processes because time pressure would, most likely, not allow to manually create a huge number of automatas for each maintenance task.

The previously described approaches mainly concentrated on the internal view of a process (e.g., are the messages exchanged in the correct order). However, a process can also be integrated by external partners. Hence, the external view must be correctly mapped to the internals. [S89] addresses this challenge with WebMov, a framework based on Modelio which is capable of modeling/generating BPEL processes including interface definitions. WebMov utilizes ActiveBPEL and BPELUnit to check if a process conforms to its WSDL model (e.g., provided operations or response times). Hence, in regard to testing the authors only provide minor extensions to existing work (i.e., how to integrate timings/time). Nevertheless it provides a very detailed evaluation which incorporates multiple case studies.

[S125] instead converts the process into an Open Nets model which is used to generate tests. The exchanged test messages are observed and compared to the public process interface. This approach mainly concentrates on optimized test-case generation, for example, by limiting test redundencies. However, it only supports Open Nets models and while the authors assume that it is possible to translate more common modeling languages such as BPMN into Open Nets they were not able to show or prove it. Finally, the presented testing approach does not support to verify data aspects.

Input/Output Symbolic transition systems are utilized in [S97,S98] to conduct black box conformance tests. Hence, the process gets converted into a STS to generate a SET. Each process partner is categorized in one of three groups, namely, observable (messages can be observed), controllable (messages can also be modified), and hidden (messages cannot be observed) to deal with controllability issues. The actual conformance test is conducted by stimulating (with manually defined test data) and monitoring the process and comparing the observed messages with the SET which is therefore used as a specification. Note, that [S98] is mainly ignoring test-case generation or test data and does mainly discuss how the communication between process partners can be observed and utilized for testing.

[S97] in contrast provides a much more complete approach but can, most likely, not be used for testing process models because the process models that are tested (BPEL) are also utilized as a form of specification which is only marginally extended. Hence, the approach provides the most value in scenarios were, e.g., a specification was implemented in some sort of programming language based on a specification which was modeled in BPEL. The applicability of this approach could be extended, for example, by providing ways to describe and test a more detailed specification which allows to specify important details beyond the ones that can are already be integrated into BPEL models.

[S96] also uses a STS. The process and the desired behavior are independently modeled/converted as/to STS models and then these models are compared. Because this approach has two independent models it does not suffer from the disadvantages discussed for [S97]. Additionally it is one of the rare testing approach that can integrate real world partner processes into the tests which allows to generate more realistic test scenarios and test results (e.g, for testing cross-organizational scenarios). However, we assume that existing modeling experts most likely do now know (without additional training) how to model STS models. Finally, this and the previous approaches (i.e., [S97,S98]) were not able to prove that their model transformation approach (e.g., BPEL process model into STS) is sound and that the generated STS model does lead to similar results as if the original BPEL model would have been used during testing - which probably reduces the test quality.

Another approach for conformance testing, in addition to monitoring the process's behavior/exchanged messages, is to record and analyze logfiles. Hence, [S119] proposes to record the exchanged messages and to compare them with local (describing a single process/partner) and global conformance rules (communication between multiple partners). The conformance rules are modeled as timed automatons and describe, for example, that message X must follow 10 minutes after message Y. The current approach only supports a posteriori analysis which increases the delay between step execution and response. Hence, adding real time analysis capabilities is a promising extensions to enable quick reactions during ongoing model executions.

A framework called XTEMP is proposed in [S121]. XTEMP allows to store and query XML based messages using a custom scripting language. Hence, conformance reports can be generated for monitored process models based on conformance rules which are represented by XTEMP scripts. The presented approach does only partly concentrate on conformance checking and mainly strives to generate management reports (e.g., the extracted indicators are not compared to an specification to detect derivations).

[S120] proposes to extract a basic set of conformance rules directly from process models. Subsequently, additional custom rules can be added by domain experts. The generated conformance rules are then used to analyze recorded event logs. The authors focus on the care domain. However, we assume that their approach would also be applicable in other domains. But therefore it must become easier to add custom extensions to the automatically generated conformance rules and the soundness of the automatic transformations must be proven.

We found that conformance testing allows to efficiently check if specific specifications are met by a process through testing the internal behavior or the external view provided for potential partners. However, we noticed that the proposed approaches are not able to combine the observed information of multiple process model instances or process models. Hence, if a constraint would be that, e.g., a manual approval for a specific transaction limit (e.g., $1,000) is required then a currently undetectable 'workaround' could be to split up the large transaction into smaller chunks. Hence, the existing approaches need to improve their data

aggregation and analyzation skills especially when monitoring process instance data and other process model runtime properties.

*3.3.7. Regression testing*

Regression testing focuses on 'retesting of a system or component to verify that modifications have not caused unintended effects' [39, p. 295]. However, a major part of the identified regression testing approaches concentrate on test-case prioritization. Why? Because, prioritization techniques allow to execute test suites in a more focused and more effective manner (e.g., by selecting tests which cover the modified process parts).

[S20] compares guards of the previous and the modified version of the process model to identify tests which must be deleted (they cover obsolete paths), modified (to follow the adapted paths), or created because a new path was added. However, existing tests are not prioritized. However, the proposed approach only identifies effected test-cases and other approaches must be used to create or adapt the existing paths. Especially adapting paths is an interesting area which is currently not addressed by existing work because mostly the tests are simply removed, therefore manual test adaptations are lost, and new tests must be generated.

Additionally, [S19] proposes to check all process partners for modifications. Afterwards only those regression tests are executed that cover at least one modified partner. [S19] also recognized that the list of chosen tests maybe must be adapted dynamically because of partner modifications which take place during test runs. Additionally multiple challenges related to concurrent process model execution paths are discussed and solved. To detect changed partners the whole partner process (or even the sourcecode of the partners implementation) must be available. We assume that this is rather unlikely in real world scenarios.

[S11] recommends three different strategies to address dynamically changing test situation. For example, it can be checked if each test follows the expected execution path. If not then most likely a partner has changed and, therefore, returned a different value. Hence, it's necessary to identify tests which likely cover the missed sub-execution paths. The authors show the efficiency of their approach by conducting a comprehensive evaluation. In [S15–S18] the authors are proposing approaches that combine multiple information pieces (the process model, WSDL interface, and XPATH rules) to calculate a test-case prioritization.

So, for the process model the branch coverage is incorporated, while the WSDL interface is analyzed to see which test covers, for example, the most operations or utilizes the most data fields [S18]; tests which use data with a higher diversity will also get assigned a higher ranking [S16]. Finally, the XPATH rules are converted into a XRG graph to apply branch coverage algorithms which allow to determine how many 'XPATH branches' are covered by a test-case [S15]. [S13,S14] create a ranking based on modified partner bindings, control flow modifications (e.g., new activities), and modified attributes (hash values of the bindings/attributes are compared between the old and new process model). Only tests whose execution path covers a change are executed.

[S13,S14] and [S12] convert processes into graphs to apply their algorithms. However, [S12] utilizes a different prioritization technique which is based on assigning a modification factor to each modified activity and its surroundings. Subsequently, the tests are ranked by adding up the covered modification factors; a higher sum then leads to a higher ranking. Another prioritization approach, based on aspect oriented programming techniques, is proposed in [S117]. Hence, the process is converted into a Petri Net to automatically weave in elements that detect changes (e.g., modified partner bindings). Overall this allows to detect which parts of the process were changed and how. Subsequently, tests are prioritized that cover those parts.

However, in order to prioritize test-cases, these must first be generated. Hence, [S14] proposes a regression test generation approach that converts the process into a graph and then extracts all possible execution paths. We found that [S14] proposes to solve constraints (e.g., defined as gateways) at each execution path to generate test input data which would allow test-cases to follow the expected execution path. But, the authors did not describe how they plan to solve those constraints and therefore their test generation approach is only partly 'operational'. [S118] instead proposes to first analyze a process to identify modified areas (using an extended annotated service automaton) and then to adapt a test suite so that it covers those areas. Note, that each process partner must provide a detailed and complex specification of its internal and expected behavior and communication. We assume that such a detailed specification will, most likely, not be available in real world scenarios.

This section provides an overview about current regression testing techniques. Clearly, test generation is not a dominating factor. However, it's possible to reuse already presented test generation approaches (cf. Section 3.3.2). Additionally we found that existing work (i.e., the majority of the identified regression testing work is concentration on test-case selection and prioritization) lacks flexibility regarding the supported regression test-case selection and prioritization user requirements. Hence, the test-cases can currently not be selected or ordered based on important requirements such as cost (i.e., each call to a process partner during a test can cost money) or execution time (i.e., how quickly the selected test-cases are executed). This limits the usability and efficiency of the presented approaches and creates an interesting area for future work.

*3.3.8. Formal testing*

Formal testing approaches utilize formal mathematical models or specifications such as labeled transition systems or FSAs to conduct process tests. However, the authors found that also publications without a strong focus on formal techniques are frequently integrating formal concepts (e.g., model checkers or LTL) to generate test-cases.

The discovered publications frequently rely on PROMELA. PROMELA is a formal verification language that is able to describe complex distributed scenarios (e.g., process models) which can be used by model checkers to generate tests based on LTL rules (cf. Section 3.3.2). [S7–S9,S92,S141] rely on PROMELA, however, the main difference is how each publication creates the analyzed PROMELA models. Hence, [S92] has designed a custom intermediate language which allows to apply their test approach on different modeling languages. So, the models are transformed into the intermediate language which is then automatically transformed to PROMELA. However, the described transformation approach is incomplete (i.e., the whole BPEL standard is not supported) and, as also already mentioned at previously discussed publications, the soundness of the transformation approach was not proven (this also applies to the following publications mentioned in this section). [S141], however, proposes to transform the original process model into an extended finite-state automaton – which is then converted into a PROMELA model – and to apply *predicate abstraction* that converts conditions into boolean states to simplify the models. Note, that this approach is one of the most complete ones (regarding the supported process model properties) and which is especially dealing with potential state explosions that can occur when using SPIN on large and complex process models.

Additional approaches based on transforming BPEL into PROMELA are presented in [S8,S9] and [S7] (the last one mainly concentrates on creating a formal, but still incomplete, representation of BPEL models and only contains a short 'discussion' how such formal models could be used for testing purposes). Additionally, [S8] lacks a throughly evaluation and only contains a performance review of the presented approach. Note that [S7] also supports our assumption that it is important to prove the equality of the original process and its formal representation in terms of their semantical meaning. Additionally [S7] stated that this prove is hard and we assume that this is the reason why the identified transformation based approaches lack this prove and frequently only support a subset of the transformed modeling language (e.g., BPEL) - which reduces the efficiency of the affected approaches, e.g., because complex real life process models likely contain some of the modeling language elements that can not yet be transformed.

Process algebra, for example, based on LOTOS which is '*one of the most expressive process algebra*' [S6, p. 224] is also frequently applied. For example, [S6] proposes to convert a BPEL process into LOTOS to test for specific properties such as X must not be followed by Y. However, it still must be defined how XPath rules can be converted. Note, that the authors mainly concentrate on the BPEL to LOTOS transformation and, therefore, skip most aspects which are required to create a complete testing framework. This gab is closed by the following publications.

[S31], uses LOTOS and LTL to create Test and Testing Control Notation behavior trees that can get assigned different types of properties which describe the expected behavior. Subsequently, counter example based test-case generation is utilized. Note, that this approach requires a very detailed and through specification (it must be created in a manual time intensive and therefore expensive way) which is then transformed into LTL rules. This approach, from all LOTOS based ones, is the only one which deals with potential state explosions. Therefore the tester has to manually identify critical process parts which are then tested while ignoring the remains of the process model. Hence,

this approach can, by applying this technique, most likely also deal with very complex and large processes but requires a modeling expert which brings in a detailed knowledge about the process model and its domain to identify the relevant critical areas.

Different process algebras are used in [S3] ($\pi$ calculus is analyzed with the Open Process Analyzer which is able to check if specific properties are met by analyzing the possible process states) and [S2] (event calculus, static analysis is used to check that execution order/data constraints (e.g., variable X must be below Y) are respected by a process model during its execution). [S3] only supports static analysis of business process models. Hence, a process model is analyzes if some possible execution order exists that should, based on the defined rules, not be possible. [S2], in contrast, supports also runtime analyses. Hence, the ongoing execution of web service compositions (i.e., comparable to process models) can be analyzed to check if it is compatible to the defined specification/rules. We assume that this would also provide an promising basis to extend the application range of this approach, for example, to adapt it so that it can also be used for security monitoring. Live monitoring active executions would also allow to deal with large process models more easily because instead of simulation and checking all possible executions only the current one must be considered and which reduces the amount of data which must be analyzed at once.

Other formal concepts are also applied. Hence, [S5] proposes to convert processes into OWL-S ontology models which are then analyzed using BLAST to generate tests based on the reachability of manually added error states. BLAST uses a C-like language as input data which must manually be extended and modified (e.g., by adding checks for variable values and states). Hence, it would support also complex checks and rules, however, we assume that a typical modeling expert would be overwhelmed by the complexity without an extensive amount of training. Additionally this approach struggles (i.e., state explosion) when dealing with complex process models especially when the process model contains concurrent paths which, we assume, as common in todays process models. [S158] however, proposes to manually create a formal representation of the process model using Algebraic Petri Nets. In the following, first order logic is used to define properties (e.g., reachability, so test-case generation is not supported) which must hold in every possible state of the process. Algebraic clustering (calculating states of specific parts of the net separately) or net unfolding (assigning fixed values to variables to reduce possible variable values and therefore states) are applied to reduce the amount of states which must be checked. This approach relies on a advanced model checker and is therefore able to deal with larger process models than already discussed approaches and it supports separation of concerns, i.e., that all test data is separately managed and stored from the original process model under test. However, this narrow focus also leads to multiple drawbacks. For example, it only supports one specific flavor of Petri Net models as input data which must be manually created based on the process model under test. This limits the applicability of this approach because competing modeling languages, e.g. BPMN, are more frequently in use nowadays. Additionally the tested properties must be defined as First Order Logic rules which likely requires that the modeling experts must be extensively trained so that they can correctly define the required rules. We assume that already those two drawbacks reduce the applicability of the presented approach in real life scenarios.

Formal testing approaches benefit from their strict mathematical definition and that existing analysis tools and theoretical background can be reused. A formal description also helps to address nondeterministic modeling languages. However, those approaches typically require that the process itself is converted into a formal representation, risking that these models (original process vs. formal representation) deviate from each other. Additionally, the presented approaches rely on modeling and specification languages which, as we assume, are most likely not known by current process modeling experts so that an extensive amount of training would be required to apply them (e.g., this hinders the dissemination of the affected process model testing approaches). It was also found that the existing formal approaches require extremely detailed specifications to create LTL rules which are used, e.g., for test-case generation purposes. However, we assume that typical project specifications or requirement documents are not detailed enough and must be substantially extended by domain and modeling experts before this approach can be applied – which is time and cost intensive.

*3.3.9. Additional approaches*

This section does not concentrate on a single topic. Instead if provides an overview about niches which are addressed by a small amount of publications.

One is to mutate a process and then to use the mutants to evaluate test suites by executing their tests and checking if the mutation is recognized (tests are failing) or not (cf. [S93–S95]). Here a test suite or testing approach would be 'better' if it finds more faults (i.e., mutations) than the compared test approaches. However, we found that the existing mutation approaches are only dealing with BPEL process models and it would be interesting which of the presented approaches can also be applied (e.g., in adapted forms) on other modeling languages such as BPMN. A through evaluation of existing BPEL mutation approaches is presented in [S94]. Other authors apply or calculate the optimal position of sensors which then communicate with a testing service to determine the test coverage of test suites, for example, at hidden private processes (cf. [S86–S88]). However, we assume that it will most likely not be possible in real world scenarios to edit and annotate the private process models of your partners with custom elements and sensors (e.g., because the partners want to protect their confidential business processes from external influences).

Sensors can also be used for testing. Thus, they have to forward relevant data (e.g., exceptions) [S85]. One disadvantage is that the process model gets annotated with additional elements which can lead to a diversion from the expected behavior and also the model becomes more complex and therefore harder to maintain. Hence, a 'modification-free' inspection is helpful which can be achieved, for example, by a proxy server which observes and analyses the exchanged messages (cf. [S82–S84]).

An alternative could be a public interface where the process partner publishes events that indicate, for example, which branches are currently covered so that an external tester can make conjectures how the test suite must be adapted to cover the whole partner process [S77]. In general, test coverage is a frequently used test suite quality indicator. Hence, [S76] proposed 4 ways of measuring the test coverage of Petri Nets. Another indicator, proposed in [S75], is to extract data invariants and to check if the existing tests are covering those. Such invariants can also be used for slicing [S73] (i.e., to identify which process paths/elements modify or are controlled by, e.g., a variable). Note, that existing approaches are mainly overtaking and adapting coverage calculation approaches which are already known from the software development domain, such as branch coverage, def-use, and so on. This might unnecessary limit the ways how coverage is observed and handled in the process modeling domain. Hence, it would be promising to analyze if the differences between source code and process models (cf. Section 3.2) are large enough to justify and successfully use novel coverage measurement approaches which are independent from the software development domain.

In general, it's also interesting to identify which activity must be fixed to resolve an identified problem. Therefore, [S110,S144] display the user the outcome of multiple executions and the user marks them as right or wrong. Those markings are then used to determine the activities which likely lead to a wrong outcome. However, this requires a substantial amount of time and in-depth knowledge of the process model and its domain which increases testing costs. Interestingly enough [S110] arguments against realistically testing cross-organizational scenarios with real process partners because those are not 'controllable' and should always be replaced with simulations. However, we assume that this argument is invalid when taking current flexible and dynamic partner and service landscapes into account. For example, process partners and their back-end systems can change and fail any time, frequently unnoticed until some processes must be executed for an important customer. Testing the process models with realistic real world scenarios, data, and the real process partner systems would be able to detect and address this situation early and increase the value of the generated test result.

An agent based testing approach is proposed in [S114]. Multiple agents simultaneously stimulate/observe a process and apply rules to validate the exchanged data. However, large parts (even basic components such as test generation) of the proposed concept were not implemented or evaluation. An extensive test framework is provided by TASSA [S50] which contains solutions for partner isolation, random data generation, test execution and definition (control and data flow based) along with a fault injection component. Note, that TASSA is, based on your assumption and findings, currently the most complete and major testing framework (e.g., a consistent user interface is provided or multiple testing approaches and solutions are combined) which we also would recommend for real world projects. Note, that other publications might provide more specialized and also more advanced solutions (e.g., it is missing the support for concurrent or cross-organizational processes which reduces its efficiency and test result quality), however, those are frequently only focusing on the scientific community and harder to apply (e.g., an extensive amount of training is required to learn custom test property description languages). Finally, [S112], shows how tests can be utilized to guide unskilled users while they are customizing process templates. However, the presented approach is only sufficient for enforcing simple control flow based rules. Hence, extending this approach, for example, to support the description and enforcement of dependences between multiple process models would be necessary to apply this approach in enterprise scenarios.

|  | Logic | Modeling | Formal techniques | Custom language | Detailed understanding | Symbolic execution | Statistics | Detailed models | Programming | Only program operation | Historical data |
|---|---|---|---|---|---|---|---|---|---|---|---|
| Conformance (21) | 38% | 67% | 48% | 52% | 71% | 5% | 0% | 19% | 0% | 10% | 0% |
| Fault testing (08) | 13% | 50% | 25% | 63% | 75% | 0% | 0% | 13% | 13% | 13% | 0% |
| Formal testing (15) | 80% | 87% | 60% | 33% | 93% | 0% | 0% | 27% | 7% | 0% | 7% |
| Integration testing (07) | 29% | 100% | 43% | 86% | 86% | 0% | 0% | 57% | 29% | 0% | 0% |
| Performance testing (04) | 25% | 75% | 50% | 25% | 75% | 0% | 0% | 75% | 25% | 25% | 0% |
| Quality of Service (09) | 44% | 56% | 33% | 33% | 78% | 0% | 11% | 33% | 11% | 11% | 0% |
| Regression testing (13) | 8% | 38% | 8% | 0% | 23% | 0% | 0% | 8% | 0% | 69% | 0% |
| Test case generation (31) | 19% | 61% | 19% | 48% | 52% | 3% | 0% | 16% | 0% | 29% | 3% |
| Test data generation (06) | 0% | 67% | 0% | 33% | 17% | 0% | 0% | 0% | 0% | 33% | 0% |
| Unit testing (14) | 7% | 79% | 7% | 50% | 57% | 0% | 0% | 0% | 7% | 7% | 0% |

Figure 6: Classification of main testing approaches and their required skills/knowledge.

It can be concluded that this section provides an overview of the state of the art in process testing, its main research areas and, therefore, found answers for RQ1 to RQ3. Additionally, it was possible to conclude that this new SLR is providing novel insights by comparing it with existing SLRs and, therefore, answering RQ0.

## 4. Classification

Classification is applied to find answers for RQ4.1 and RQ4.2 such that it becomes possible to identify key research areas, novel research challenges, and gaps in the process model testing domain.

### 4.1. Required skills/knowledge

For RQ4.1 the classification compares the discovered publications based on the dimension 'Approach' and 'Skill/Knowledge' at Figure 6 to evaluate if it's possible that a 'typical' modeling expert would be able to apply current testing techniques. The vertical axis contains the identified main testing approaches (cf. with Section 3.3) while the horizontal axis consists of the identified skills/knowledge which are required to apply those testing approaches. Skills/knowledge were divided in following classes:

**Logic** indicates that the user needs skills in logic, for example, to define test properties using LTL.

**Modeling** skills are required, i.e., a process model can be altered or generated.

**Formal techniques** are required, for example, to create a formal model that represents a specification.

**Custom language** requires that a custom test description language must be applied.

**Detailed understanding** means that the user needs in depth knowledge about the process and its environment and domain.

**Detailed models** means that the user is required to annotate the models with fine-grained details, for example, pre- and post-conditions at each activity.

**Programming** skills are required, for example, to define a simulation of a partner by describing its behavior in Java.

**Only program operation** means that it was *only* required to apply the test algorithms/tools.

**Historical data** refers to the ability to utilize historical information such as process execution log files.

Note, that the number in brackets (cf. Figure 6 and 7) indicate the number of publications that chose this specific testing approach as their main objective. The cells indicate the relative amount of publications which require, for example, a specific approach/skill combination (in percent). The shaded colors visualize the values (a higher percentage is represented by a darker color).

Figure 6 shows that the published testing approaches typically require manual preparation. However, it's possible to identify unique characteristics. 69 percent of all regression testing approaches can be applied simply by executing their algorithms. However 9 out of 10 publications (forming this 69 percent) are test-case *prioritization techniques*.

By contrast, detailed models are typically required by performance testing approaches because those utilize information about the expected execution times. Modeling skills are also frequently expected. However, these are likely available at modeling experts while the authors assume that the other skills (except program execution and historical data) are hardly available. But it must be kept in mind that Figure 6 is counting each required skill individually, meaning that only 4.7 percent of the discovered publications were *only* requiring modeling skills.

| Approach (No) | Conformance (21) | Fault testing (08) | Formal testing (15) | Integration testing (07) | Performance testing (04) | Quality of Service (09) | Regression testing (13) | Test case generation (31) | Test data generation (06) | Unit testing (14) |
|---|---|---|---|---|---|---|---|---|---|---|
| **Testing cross-organizational scenarios** | | | | | | | | | | |
| Yes | 24% | 13% | 7% | 86% | 25% | 22% | 46% | 13% | 0% | 7% |
| No | 76% | 88% | 93% | 14% | 75% | 78% | 54% | 87% | 100% | 93% |
| **Consideration of legacy scenarios** | | | | | | | | | | |
| Yes | 0% | 0% | 0% | 0% | 0% | 0% | 0% | 0% | 0% | 0% |
| No | 100% | 100% | 100% | 100% | 100% | 100% | 100% | 100% | 100% | 100% |
| **Applying model abstraction/transformation** | | | | | | | | | | |
| Yes | 71% | 25% | 93% | 71% | 50% | 44% | 38% | 68% | 0% | 29% |
| No | 29% | 75% | 7% | 29% | 50% | 56% | 62% | 32% | 100% | 71% |

Figure 7: Classification of main testing approaches and their support for upcoming challenges.

However, we also checked which approaches typically require a complex combination of skills (taking all skills into account except modeling, historical data, and program execution – we assume that those should not restrict the applicability). We found that integrating testing (each skill is required, on average, by 41 percent of the concerned publications), formal testing approaches (37 percent), and performance testing (34 percent), require the most complex skill combination and therefore provide only a limited applicability. Another noticeable circumstance is that a 'Detailed understanding' (in-depth knowledge of the process) is required frequently, for example, to describe the expected behavior when a mangled message is sent to the process (for fault testing).

We conclude that specific knowledge sources, especially historical data, are currently not exploited to their full extent and that current testing approaches frequently require a multitude of skills.

*4.2. Support for upcoming challenges*

Figure 7 compares the discovered publications based on the dimension 'Approach' with multiple, important [34], upcoming challenges to evaluate if those are already supported by existing process testing approaches. Note, that 'cross-organizational scenarios' and 'legacy scenarios' were already described in Section 1.

**Applying model abstraction/transformation:** Current research frequently proposed solutions which required that the original process model becomes transformed/abstracted (cf. Section 4.2), for example, to a STS. However, the applied transformation/abstraction techniques frequently only supported a limited set of the features (compared which original process modeling language). In addition, it's hard to ensure that the transformed result does not derivate from the original model (cf. [34] and [S7]).

Figure 7 shows that the support for cross-organizational scenarios is quite low. However, integration testing and regression testing sticks out. The reason is that integration testing tests the integration of a process with other partners/services (cf. Section 3.3.4) and regression testing naturally has to deal with mutable partners, for example, to prioritize tests which cover a modified partner process (cf. Section 3.3.7). But the support for cross-organizational scenarios would also be important for the other approaches because process models contain more and more cross-organizational scenarios.

Additionally, it was noticeable that legacy scenarios are currently not taken into account by the discovered testing approaches. Hence, the discovered work does typically require manual preparation or various skills, which limits the applicability of the proposed approaches in real live scenarios (cf. Section 1). Drawbacks and limitations of such end-user prerequisites are currently not discussed and the challenge to maintain legacy processes is not systematically addressed.

Applying model abstraction/transformation is a popular approach because it enables researchers to utlize existing tools/techniques (e.g., model checkers are used by 'Formal testing' approaches). However, its drawbacks (cf. Section 5) aren't discussed by the discovered publications. But why is 'Test data generation' not using any transformation/abstraction? Because the identified main data generation techniques (cf. Section 3.3.3) are not benefiting it.

We conclude that specific testing approaches (performance testing (4 publications), fault (8), and integration testing (7)) are only partially covered by existing publications and that those are, therefore, still emerging research areas. Specifically, legacy scenarios (and partly cross-organizational scenarios) are not addressed by current publications and we conclude that this is hindering the dissemination and applicability of tests at the process modeling domain.

## 5. Findings and research challenges

This section discusses findings, research challenges, and gaps in the process testing domain which were discovered based on this SLR and the applied classification.

*Legacy process models.* The review shows that, current process testing research does not take legacy process testing scenarios into account and expects the existence of a supportive user which has to bring in a bunch of knowledge/skills. Few of the proposed approaches can be applied without manual preparation (e.g., some test-case prioritization and generation approaches). However, even these approaches do not explicitly try to address legacy scenarios (e.g., by assuming an insufficient documentation, limited access to domain/environmental information, and a realistic training/skillset of the user). Hence, future work must take such scenarios into account to support modeling experts when they are *maintaining and refining* millions of existing *legacy process* models.

*Cross-organizational scenarios.* It is essential for testing approaches to support cross-organizational scenarios in order to unfold their full potential [34]. However, the conducted classification showed that this was typically not the case (cf. Section 4.2). But the trend towards interconnected organizations cannot be reverted. Hence, process testing research is forced to support cross-organizational scenarios in future publications to provide a high applicability in real life scenarios.

*Model transformation and abstraction.* Current testing techniques frequently concentrate on BPEL and its particular characteristics. Hence, to widen and enhance the applicability of process tests it's necessary to propose generic solutions (e.g., through intermediate languages, cf. [S92]) or approaches which address the *unique characteristics* of other modeling languages (e.g., the possible nondeterministic behavior of BPMN models). However, when applying transformation/abstraction (between modeling languages or into intermediate languages) it must be ensured that the tests still check the original behavior – a challenge which is currently frequently ignored (cf. Section 4.2).

*Skills and knowledge.* Current test approaches require a multitude of skills/ information. However, we could not discover a publication which discusses the implications of such requirements (e.g., that not every user would be able to apply them without extensive training or that the applicability of tests is limited in specific scenarios). Such a scenario is the maintenance/advancement of legacy processes. We assume that in such a scenario time pressure and frequently changing domains/tasks are immanent.

Taking into account the complexity of current process models and domains it's unlikely that – in a reasonable amount of time – a user could gain enough knowledge to describe, for example, detailed conditions for each activity. Therefore, research should strive to reduce the amount of skills which are required to apply the proposed testing approaches and analyze which skills can be assumed as being existent at a typical user of process tests.

*Integrating historical data.* Testing research rarely utilizes historical data sources, for example, to forecast execution times (cf. [S23]). However, historical data can also be utilized to enhance the quality of the generated test-cases and to reduce the amount of information which must be provided manually. For example, historical data could be used to a) detect the most frequently executed and therefore probably most important paths, b) apply long time monitoring to detect derivations from a auto generated behavior fingerprint, and c) to auto-generate complex test data based on the recorded information.

*Comprehensive account of process views.* A process model typically contains different views which focus on various properties such as control flow, data flow, resources, and organizational properties. However, the current process testing research mainly concentrates on the first two ones and leaves the resource and organizational view behind. Hence, the questions arise, if and how those views can be covered and utilized by test approaches.

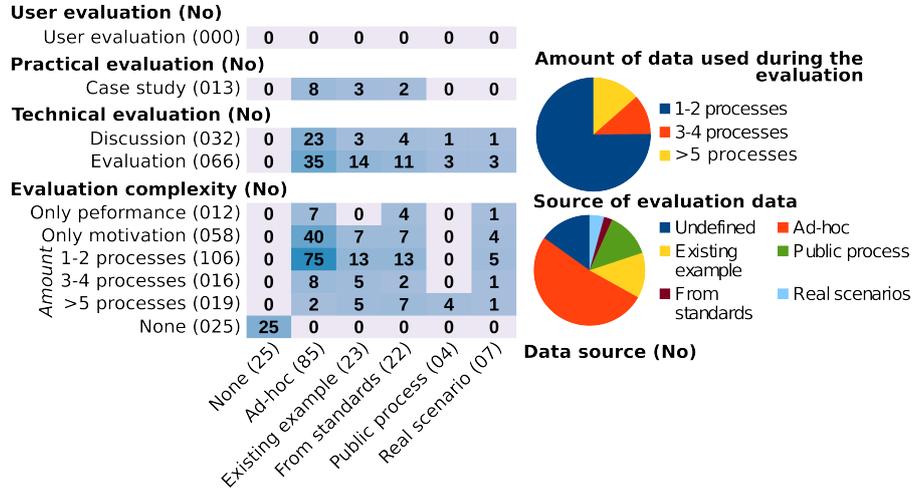

Figure 8: Overview of the discovered evaluation approaches.

*Applied evaluation approaches.* Figure 8 provides an overview of the evaluation techniques which were applied by the discovered process testing publications. Therefore it structures the Y axis into three evaluation approaches concentrating on the *user*, the *practical applicability* of the proposed approaches and *technical aspects*. The 'Evaluation complexity' section provides a quantitative overview of the utilized evaluation data (e.g., 'Only motivation': only a motivational example was provided, 'Only performance': only a performance evaluation was conducted, or 'Amount': how many process models were used during the evaluation). The X axis shows the source of the utilized evaluation data, for example, 'Ad-hoc' indicates that the publication used ad-hoc process models or 'Real scenario' indicates that some real life process model was used that is not publicly available. 'None' indicates that no evaluation was presented.

It can be deducted from this figure that current research is frequently not providing a comprehensive evaluation (16.3 percent not even provided an evaluation or motivational example). Instead a typical process testing publication only used a limited amount (1-2) of processes (49.6 percent) to evaluate an approach, only provided a motivational example (26.1 percent), or a performance evaluation (four percent). The performance evaluations were normally only 'proving' that the proposed techniques can be applied on X processes in Y minutes without comparing these to different algorithms. We also found some publications whose authors discussed opinions/experiences instead of conducting a full-blown evaluation. The evaluations itself frequently relied on ad-hoc process models which the authors have designed specifically for their evaluation section. However, there was also a noticeable amount of publications which re-used existing examples (e.g., from a different publication) or standardized models which allows to compare different approaches. One solution could be to create a platform were process models, test-cases, and evaluation results can be shared between authors.

It was also noticeable that current process testing approaches are *not* conducting user evaluations (e.g., interviews or questionnaires). Hence, the usability of the proposed approaches in real life scenarios is often unproven. One step in the right direction would be to conduct more case studies. However, the discovered case studies are typically using ad-hoc process models. This finding was surprising, because we expected that case studies should be based on real live scenarios.

*Misuse-cases.* It became obvious that a considerable amount of the discovered publications was focusing on applying tests to check if a process supports a specific property or use-case (e.g., is the expected message returned after sending X). So, a typical process testing publication checks if some *expected* behavior is supported. But, it would also be interesting to ensure the non-existence of specific behavior (e.g., to prevent misuse-cases or to guarantee with tests that a fixed unintended *feature* is not 'resurrecting' in future versions).

Overall it can be concluded that process testing is a challenging research field. However, this listing does not claim to be complete and therefore we assume that new challenges will arise.

## 6. Discussion of limitations and threats to validity

This section discuss limitations and potential threats to validity of this SLR.

*Excluded areas.* This SLR concentrated on discovering and analyzing process model testing publications. However, while designing this SLR the authors chose to a) ignore *security testing* approaches because there was a recent publication of a SLR in that area (cf. [6]) and b) of approaches which are not testing the process models itself but instead some source code based representation to check the conformance between code and model. [52]).

*Limitation of search.* Due to the wide-spread utilization of processes it became necessary to limit the search (cf. Section 2.2) to specific sections of the searched publications (title, abstract, and keywords). For example, a full text based search after *process* and *test* generates more than 220,000 results on Google Scholar alone. This decision also slightly limited the possible search engines because, for example, Google Scholar does not allow to specifically search the abstract of the indexed publications. However, this SLR still includes more than seven different literature databases, conducted a detailed manual search, and applied snowballing. Therefore, we are confident that this SLR covers the most important sources and publications.

*Presented ideas and algorithms.* This SLR elaborates the ideas, algorithms, and foundations of the discovered publications. However, an extensive elaboration would go beyond the scope of this study. Hence, we would like to refer the interested reader to the cited literature.

*Manual categorization.* A manual categorization always struggles with the inherent problem of a potential bias. Hence, we discussed each publication until an agreement was reached and investigated automatic categorization tools. However, we found that text mining is not suitable for our problem (cf. [6]) because the wording changed significantly over the years and the classified information was frequently not described using extractable keywords.

*Classification.* We are aware that the classification section can (for the sake of brevity) only present some selected examples to generate an overview and to identify research challenges. However, we also created additional analyzes and found that those were **not** contradicting the presented results.

## 7. Conclusion and future work

This SLR provides a holistic view on process model testing. Therefore, 153 publications between 2002 and 2013 were examined, analyzed, and classified. Section 5). Related to our research questions we found that:

**RQ0** We became confident – by searching and analyzing existing SLRs – that this SLR is the most complete SLR on the process testing topic and that it covers novel research questions (cf. Section 3.1.1).

**RQ1** This SLR indicates that existing research is interconnected with the software development domain. Therefore, the community is reusing existing definitions which have already proven their usefulness. However, the most recent publications suggest that the process testing community is currently on its way to abstract itself from its roots by publishing more unique approaches.

**RQ2** We found a rich set of different approaches which address ten main objectives (cf. Section 3.3) and that the majority of the discovered publications can be mapped onto one of those. However, there were also some distinct publications which address objectives such as mutation testing.

**RQ3** By subdividing the discovered publications based on their main objective it became clear that the most frequently addressed topic is test-case definition with 31 publications (cf. Section 3.3). Note, that additional publications could also implicitly cover this topic; but those are not counted at this category if test-case generation is not their main objective. Conformance testing also got a top position assigned with 21 publications. However, the least frequently addressed topic is performance testing with four publications.

**RQ4** In terms of RQ4 a mixed picture emerges. It was found that current process testing approaches typically require a multitude of skills and knowledge, while at the same time auto generated knowledge sources, such as historical data, were not fully exploited. Additionally, an encouraging trend towards the support for cross-organizational scenarios and test approaches which do not require model abstraction were recognizable. However, further improvements could still be made, by creating process testing approaches which keep legacy scenarios in mind. We conclude that current process testing approaches are on the right track to address upcoming challenges, but there are still many points remaining were a substantial improvement is possible (cf. Section 5).

In future work, we plan to strengthen the process model testing domain by focusing on the unique characteristics and requirements of the process modeling community (e.g., by addressing modeling languages which are less strongly rooted at the software development domain). Further, we want to tackle open research challenges and problems which were outlined in this paper. Therefore, we will, for example, concentrate on developing process testing approaches which utilize the full potential of historical data and comparable data sources which are auto generated by process aware information systems.

## References

Reference numbers which are prefixed with a S are available at the appendix.

### Appendix

This supplementary file contains a complete list of the manually searched conferences and journals used in the article Kristof Böhmer and Stefanie Rinderle-Ma: A systematic literature review on process testing: Approaches, challenges, and research directions (2015) submitted to the Information and Software Technology journal. Additionally, it contains a list of all process test publications and relevant existing SLRs which were discovered during the SLR.

### Appendix A. Complete List of manually searched journals and conferences

Business Process Management Journal (BPM), Central European Journal of Computer Science (CEJCS), Computers in Industry (COMPI), The Computer Journal (COMPJ), Computer Science and Information Systems (CSIS), International Journal of Asian Business and Information Management (IJABIM), International Journal of Business Information Systems (IJBIS), International Journal of Business Process Integration and Management (IJBPIM), International Journal of Cooperative Information Systems (IJCIS), International Journal of Information System Modeling and Design (IJISMD), International Journal of Knowledge-Based Organizations (IJKBO), Information Systems Management (ISM), MIS Quarterly (MISQ), IEEE Transactions on Knowledge and Data Engineering (TKDE), *Information Systems (IS)*, *Data & Knowledge Engineering* (DKE), *IEEE Transactions on Software Engineering* (TSE). The following conferences were searched: International Conference on Application and Theory of Petri Nets and Concurrency (ICATPN/APN), International Conference on Business Process Management (BPM), Conference on Advanced Information System Engineering (CaiSE), IEEE Conference on Business Informatics (CBI), International Conference on Cooperative Information Systems (CoopIS), OTM Conferences CoopIS DOA-Trusted Cloud and ODBASE (OTM), European Conference on Information Systems (ECIS), Business Process Management with Event-Driven Process Chains (EPK), IEEE International Conference on e-Business Engineering (ICEBE), International Conference on Tests and Proof (TAP), International Conference on Testing Communicating Systems (TestCom), *IEEE International Conference on Web Services* (ICWS), *International Conference on Service Oriented Computing* (ICSOC).

The journals and conferences which are written in italic were added to the list during the discussions on the preliminary protocol, all other sources were already included from the beginning. The journals and conferences were selected based on their relevance and impact factor (Journal Citation Reports 2011 and Microsoft Research conference field raking).

### Appendix B. References containing the discovered process test publications and existing relevant SLRs

[S1] Z. J. Li, S. Wei, BPEL-Unit: JUnit for BPEL Processes, in: Service Oriented Computing, Springer, 2006, pp. 415–426.